\documentclass[a4paper, onecolumn]{jov_arxiv}

\usepackage{amsfonts,amsmath,amssymb}
\usepackage{mathrsfs}
\usepackage[utf8]{inputenc}
\usepackage[T1]{fontenc}
\usepackage{mathbbol}
\usepackage{natbib}
\usepackage{color}
\usepackage{url}

\usepackage{dsfont}
\usepackage{cancel}
\usepackage{pbox}
\usepackage{bbm}

\usepackage[left]{lineno}
\usepackage{array}

\usepackage{rotating}
\newcommand{\red}[1]{\textcolor[rgb]{0,0,0}{#1}}
\newcommand{\blue}[1]{\textcolor[rgb]{0,0,0}{#1}}
\newcommand{\green}[1]{\textcolor[rgb]{0,0,0}{#1}}

\newlength\savedwidth

\usepackage[all]{xy}

\newcommand{\vect}[1]{\boldsymbol{#1}}
\newcommand{\Felix}[1]{ {\bf \textcolor{red} {Felix: #1} }}

\begin{document}
\setpagewiselinenumbers
\modulolinenumbers[1]



\title{Estimating the contribution of early and late noise \\[0.3cm] in vision from psychophysical data}

\abstract{
Human performance in psychophysical detection and discrimination tasks is limited by \emph{inner} noise. It is unclear to what extent this inner noise arises from \emph{early} noise (e.g.\ in the photoreceptors), or from \emph{late} noise (at or immediately prior to the decision stage presumably in cortex). Very likely the behaviourally limiting inner noise is a non-trivial combination of both early and late noise. Here we propose a method to quantify the contributions of early and late noise purely from psychophysical data.
Our approach generalizes classical results for linear systems~\citep{Burgess88}  by combining the theory of noise propagation through a nonlinear network \citep{Ahumada87} with expressions to obtain a perceptual metric through a nonlinear network \citep{Malo06a,Laparra10a}. We show that from threshold-only data the relative contributions of early and late noise can only be disentangled when the experiments include substantial \emph{external} noise.
When full psychometric functions are available, early and late noise sources can be quantified even in the absence of external noise. Our psychophysical estimate of the magnitude of early noise---assuming a standard cascade of linear and nonlinear model stages---is substantially lower than the noise in cone photocurrents computed via an accurate model of retinal 
physiology, the ISETBio~\citep{Cotaris1}.
This is consistent with the idea that one of the fundamental tasks of early vision is to reduce the comparatively large retinal noise.

}

\author{Malo}{Jes\'us}
 {Image Processing Lab, Parc Cientific}
 {Universitat de València, Spain}
 {http://isp.uv.es}
 {jesus.malo@uv.es}
\author{Esteve-Taboada}{Jos\'e Juan}
 {Dept. Optics and Optometry and Vision Science, School of Physics}
 {Universitat de València, Spain}
 {http://}{josejuan.esteve@uv.es}
\author{Aguilar}{Guillermo}
 {Computational Psychology}
 {Technische Universit\"{a}t Berlin, Germany}
 {http://}{guillermo.aguilar@mail.tu-berlin.de}
\author{Maertens}{Marianne}
 {Computational Psychology}
 {Technische Universit\"{a}t Berlin, Germany}
 {http://}{marianne.maertens@tu-berlin.de}
\author{Wichmann}{Felix}
 {Neural Information Processing Group}
 {University of T\"ubingen, Germany}
 {http://}{felix.wichmann@uni-tuebingen.de }

\keywords{Early noise, late noise, external noise, spatial vision, pattern discrimination, linear-nonlinear models, information transmission.}

\maketitle

\section{1. Introduction}
\noindent One goal of psychophysical research is to relate physical to perceived stimulus properties. An important characteristic of this relationship is its variability. When the same stimulus is presented multiple times (which is common in psychophysical experiments), it might evoke different behavioral responses, at least when it is neither trivially easy nor difficult to perceive (neither ``at ceiling'' nor ``at floor''). 
Since the advent of signal detection theory (SDT) in the 1950's the source of this behavioural variability is conceptualised and modelled as arising from inner noise on a putative decision- or evidence-axis \citep{Tanner_1954,Swets_1961,Green_1988}. In this widely accepted view, perceptual detection and discrimination behaviour depends on the strength of the inner response to a stimulus relative to the inner variability (``signal and noise'').

From a neurophysiological standpoint, noise sources in sensory systems are multiple and arise at every stage within the system. In the visual system neuronal variability (noise) is observed for photon detection in the photoreceptors as well as for all neurons along the visual pathways. The effect of the different neurophysiologically measured noise sources on observable psychophysical behaviour is unclear, however.

Despite the neuroscientific identification of many noise sources, perceptual models of human visual behaviour commonly refer to all noise sources together as \emph{inner} noise \citep{Burgess88}. \blue{Often}, inner noise is modelled as a single late noise source, which is added as decision noise after completion of all processing stages in a model. The discrimination ability of a model is hence determined by the variability of the model output, i.e. after the stimuli have propagated through and are changed by the system. 
Discussion of noise in psychophysical models is predominantly limited to the type of \emph{late} noise, i.e. whether its variance is fixed or varies with signal strength~\citep{Wichmann_1999,Georgeson_2006}.  
A discussion of the number of noise sources and their position within the (nonlinear) processing stages is comparatively rare (but see \citealp{Pelli_1991,Pelli99,Wichmann02}).


We can think of at least two reasons why the number and type of noise sources in behavioural models of perception have been rarely explored and are thus still wide open:
\begin{enumerate}
    \item Psychophysically, all we can measure is the probability of correct detection or discrimination as a function of stimulus intensity or stimulus differences. In both cases performance is co-determined by non-linearities of stimulus processing and by noise, and hence inferences about the type of noise can only be made for a fixed nonlinear transformation (and vice versa).
    \item Inclusion of an early noise source complicates matters considerably, because early noise, like the stimulus itself, will be nonlinearly transformed through the system. This typically results in a noise term for which no closed-form expression exists. The behaviour of such a model would thus require (extensive) numerical simulations, and fitting such a model to psychophysical data is likely to be cumbersome.
\end{enumerate}

Here we propose a strategy to separately estimate at least two different components of noise in the visual system, \emph{early} and \emph{late}. Roughly speaking we attribute early noise to photoreceptors and late noise to all subsequent neuronal processes.
We use \emph{external} noise, i.e. noise applied to the external stimulus, whose magnitude and properties are under the complete control of the experimenter, to estimate early and late noise components from psychophysical data. We adopt ideas from the theory of noise propagation through a network \citep{Ahumada87} 
and the transforms of distance metrics in a network \citep{Malo99,Malo06a,Laparra10a}
to derive a perceptual metric that depends on the different noise sources and is calibrated by the external noise.

\subsection{Outline}
\noindent The structure of the paper is as follows: 
In section~2 we introduce our modeling framework, the notation and a simplified vision model which illustrates our method. We also describe the experimental data that we use to estimate early and late noise parameters.
In section~3 we propose a method to estimate the noise sources using threshold-only data. The idea is to propagate a stimulus perturbed by multivariate noise through a vision model and to compute a noise-dependent Mahalanobis distance metric from these multiple propagations. We show how our method generalizes the classical result of~\citep{Burgess88}, and discuss the key role of external noise to determine the relative magnitude of early and late noise sources in threshold-only experiments. 
In section~4 we show how we can estimate the relative magnitude of early and late noise sources even without external noise when we have the full psychometric function and not just thresholds. 
Section~5 discusses the connection of our method with physiological estimates of (early) retinal noise, and with other psychophysical methods for (inner) noise estimation. We also outline the implications of our estimates for other experimental methods and for information-theory approaches to study vision.
Appendices present mathematical proofs, fitting procedures, specific simulation results, and suggestions how to extend the presented methods.


\section{2. Modelling framework and intuition}
\label{sec_theory-and-notation}

\subsection{2.1 Notation}
\noindent We assume a vision model $S$ which transforms the visual stimulus, the vector $\vect{x}$, into a representation of the stimulus, the vector $\vect{y}$. We assume the transform $S$ to be deterministic. The input to $S$ is assumed to be a mixture of the stimulus $\vect{x}$ and {\bf\emph{early noise}} $\vect{n}_e$, which is added to $\vect{x}$ at an early stage, presumably the retina. 
We add {\bf\emph{late noise}} $\vect{n}_l$ to the output from $S$, which subsumes and reflects the noise that may happen at various levels of visual processing summarized in $S$. 
Early and late noise lead to a single noise term {\bf\emph{inner noise}} $\vect{n}_\mathcal{I}$, which is added to the deterministic response of the system. The {\bf\emph{inner noise}}, $\vect{n}_\mathcal{I}$ is the difference between the noisy response and the deterministic response to the stimulus, $\vect{n}_\mathcal{I} = \vect{y} - S(\vect{x}) \label{inner_noise}$. These concepts are summarized in the following diagram:
\vspace{-0.1cm}
\begin{eqnarray}
  \xymatrixcolsep{2pc}
  \xymatrix{
     \vect{x} + \vect{n}_e
  \,\,\,\,\,\,\, \ar@/^2.3pc/[r]^{\scalebox{1.0}{$S$}} & 
  \,\,\,\,\,\,\,\,\,\,\,\,\,\,\,\,\,\,\,\,\, \vect{y} = S(\vect{x} + \vect{n}_e) + \vect{n}_l
  }
 \label{noisy_resp} \\
  & \!\!\!\!\!\!\!\!\!\!\!\!\!\!\!\!\!\!\!\!\!\!\!\!\!\!\!\!\!\!\!\!\!\!\!\!\!\!\!\!\!\!\!\!\!\!\!\!\!\!\!\!\!\!\!\!\!\!\!\!\!\!\!\!\!\!\!\!\!\!\!\!\!\!\!\!\!\!\!\!
  \vect{y} = S(\vect{x}) + \vect{n}_\mathcal{I} \nonumber
\end{eqnarray}

It is the noise \textit{after} the transformation $S$  (inner noise, $\vect{n}_\mathcal{I}$) which limits the performance of an observer when discriminating two stimuli $\vect{x_1}$ and $\vect{x_2}$, in particular when the difference of system responses $\Delta S = S(\vect{x_2}) - S(\vect{x_1})$ is small relative to the inner noise, $\vect{n}_\mathcal{I} \gg \Delta S$. 
We call the first stage, \textit{before} $S$ has been applied, `early representation', and the second stage, \textit{after} $S$ has been applied, `late representation'.


\subsection{2.2 Vision model}

\noindent The vision model we adopt as the deterministic transform $S$ is a standard \emph{linear+nonlinear} cascade. The model consists of a linear stage with wavelet filters which are scaled in agreement with the contrast sensitivity function (CSF; \cite{Robson_1966}), followed by a point-wise nonlinearity $m$, which represents a simple model of masking:
\begin{equation}
  \xymatrixcolsep{2pc}
  \xymatrix{ \vect{x} + \vect{n}_e \,\,\,\,\,\,\,\,\,\,\, \ar@/_1.5pc/[r]_{\scalebox{1.05}{$W \equiv \textrm{wavelets}$}} \ar@/^2.3pc/[rrr]^{\scalebox{1.05}{$S$}} &  \vect{w}  \ar@/_1.5pc/[r]_{ \scalebox{1.05}{$\mathbb{D}_{\mathit{CSF}}$} }  & \,\,\,\,\,\,\,\, \vect{w'} \ar@/_1.5pc/[r]_{\scalebox{1.05}{$m(\vect{w}') \equiv \textrm{nonlinearity}$}} & \,\,\,\,\,\,\,\,\,\,\,\,\,\,\,\,\,\,\,\,\,\,\,\,
  \vect{y} = S(\vect{x} + \vect{n}_e) + \vect{n}_l
  }
  \label{modular}
\end{equation}
 
\noindent The input to the deterministic model $S$ consists of the stimulus plus the early noise, $\vect{x} +\vect{n}_e$, expressed in units of $cd/m^2$.
Our estimation methods, as seen in Eqs.~\ref{metric_noise1} and~\ref{psychometric} below, are general in the sense that they are not attached to a specific noise formulation, may depend on the input~\citep{Cotaris1,Cotaris2,Dayan01} and include correlations~\citep{MorenoBote14,Pouget16}. For the sake of simplicity, here we assume that the early noise follows a simple Gaussian-Poisson distribution~\citep{Wichmann_1999,Cotaris1,Cotaris2,Dayan01} (see Eq.~\ref{Gaussian-PoissonCOV} in Appendix A for the expression of the signal-dependent covariance). The psychophysical scaling parameter (Fano factor, $\beta_e$) of this distribution is unknown and its determination is one of the goals of our method.

\begin{figure}[t!]
\begin{center}
   \includegraphics[width=1.0\textwidth]{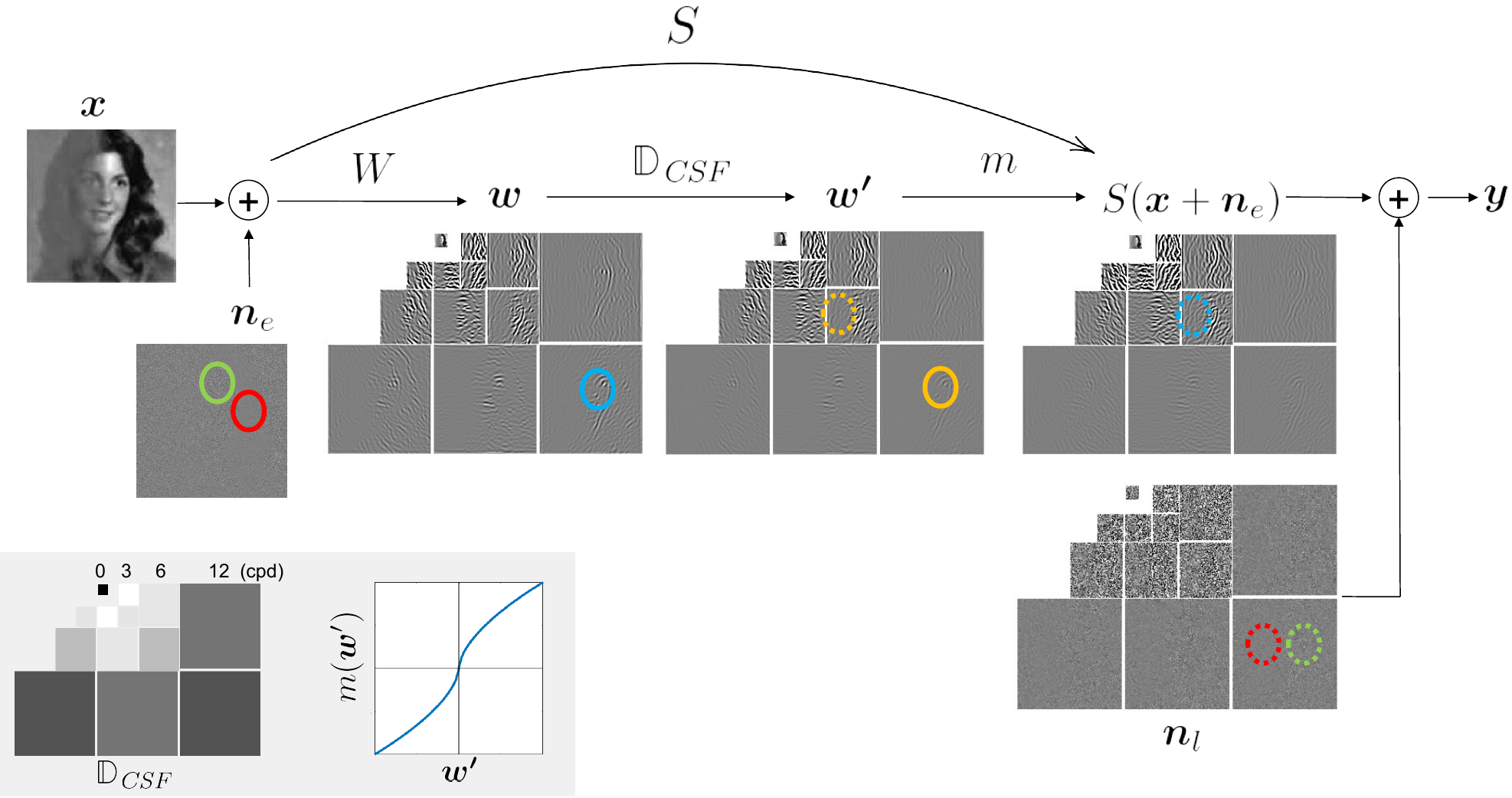}\\
   \caption{\blue{\textbf{Illustration of the vision model.} The input image, $\vect{x}$, is corrupted by early noise, $\vect{n_e}$, leading to an \textit{early (noisy) representation}. 
   In this example we assume Poisson noise and its amplitude increases with the luminance of the input (see the noise in the green circle with respect to the noise in the red circle). 
   Then the signal is analyzed by a set of linear wavelet-like oriented filters (tuned to 0, 3, 6, 12 and 24 cpd), with responses $\vect{w}$. 
   In the figure we use the classical representation of subbands in the wavelet literature~\citep{Simoncelli92}. The 24 cpd subband is not represented for clarity.
   Then, the responses are multiplied by the CSF weights in the wavelet domain~\citep{Malo97} ($\mathbb{D}_{\mathit{CSF}}$), shown in the lower left inset. Here the lighter gray correspond to bigger weights. The represented CSF shows the band-pass behavior and the oblique effect.
   CSF weighting is apparent in the attenuation of high-frequency subbands in $\vect{w'}$: see how the  energy in responses in the solid blue circle is reduced in the solid yellow circle. Then, the responses undergo a fixed saturating nonlinearity, $m(\cdot)$~\citep{Legge81}, that preserves the relative scale of each subband~\citep{Martinez19,Malo24}. This nonlinearity takes the low amplitude responses (e.g. in the dashed yellow circle) and leads to enhanced responses (in the dashed blue circle).
   Finally, late noise, $\vect{n_l}$, is added to the responses in this \textit{late representation}.
   The Poisson nature of the late noise is apparent in that its amplitude in all subbands is larger in the spatial region with larger contrast and hence larger response of texture detectors (see larger noise at the right, e.g. in dashed green circle, and less noise at the left, dashed red circle).
   The amplitude of the early and late noises has been scaled for clarity ($\times 40$ and $\times 4$, respectively).}}
\label{full_model}
\end{center}
\end{figure}

The first step in the model $S$ is a set of local, oriented filters at different scales~\citep[see, e.g.,][]{Watson87b}. This linear transform is followed by a scale dependent weighting which simulates the effect of the CSF. We use a steerable wavelet transform~\citep{Simoncelli92} for the linear filters and the method detailed in~\citet{Malo97} to obtain the optimal CSF weights in that domain. This linear stage can be summarized as the product of two matrices: $\vect{w}' = \mathbb{D}_{\mathrm{CSF}}\cdot W \cdot (\vect{x} + \vect{n}_e)$, where $W$ contains the wavelet receptive fields in rows and $\mathbb{D}_{\mathrm{CSF}}$ is a diagonal matrix with weights that represent the CSF in the diagonal.
In the final step we apply point-wise saturating functions to each coefficient of the linear transform~\citep{Nachmias_1974,Legge80,Legge81}.
Specifically, $\vect{y} = m(\vect{w}') + \vect{n}_l = K \odot \mathrm{sign}(\vect{w'})\odot|\vect{w}'|^\gamma + \vect{n}_l$.
$\vect{a} \odot \vect{b}$ represents the element-wise product of vectors $\vect{a}$ and $\vect{b}$, the exponent $\gamma$ is applied element-wise to the absolute value of every component in $\vect{w'}$, and the original sign of the components is preserved through the $\mathrm{sign}(\cdot)$ function. 
We apply a correction at the origin to avoid singularities in the derivative (see the formulation of the $\gamma$-nonlinearity in~\citep{Martinez18,Malo24}), and we choose the constant $K$ so as to keep the relative magnitude of the sub-bands, and hence preserve the effect of the CSF~\citep{Martinez19}.

Late noise is applied to the output of the model transform $S$. Again for convenience we assume that the late noise $\vect{n}_l$ follows a Gaussian-Poisson distribution (see covariance in Eq.~\ref{Gaussian-PoissonCOV} in Appendix A), and the goal is to obtain the Fano factor, $\beta_l$, of this distribution. We assume level-dependent early and late noise sources for explicit exposition in this work. However, our method should also allow to investigate the role of level-independent late noise in models of spatial vision \citep[see e.g.][]{Wichmann_1999,Kontsevich_2002,Georgeson_2006,Wichmann17}, an issue we return to in the discussion.

\blue{A step-by-step visual illustration of the transforms in the model is shown in Fig.~\ref{full_model}.
A Matlab implementation of the model, the description of its parameters, and the data and methods used to estimate the noise levels is available online\footnote{\texttt{http://isp.uv.es/code/visioncolor/noise.html}}.}
The model is somewhat simplistic, and surely not state-of-the-art in early spatial vision. Its elements, the wavelet transform and the CSF, are taken off-the-shelf and we chose reasonable values for its free parameters, the saturation exponent, $\gamma$, and the sub-band-dependent constant, $K$. 
However, this \emph{reasonably good} and \emph{reasonably simple} model serves our present purpose to illustrate the method of noise propagation to derive a perceptual metric. 
It is reasonably good, because it allows us to predict human data from detection and discrimination experiments with periodic stimuli (shown below). It is reasonably simple, because it has well behaved derivatives, $\nabla S$, for computations in the optimization. In principle our method can be applied to more sophisticated models of the transform $S$. However, the estimation of the derivatives might become more cumbersome, or may require numerical approximations.

\subsection{2.3 Model response in the presence of noise: a two-pixel example}
\noindent To provide an intuition for the effects of early vs. late noise on the representation of the stimulus, we look at the model responses in a toy example of two-pixel images. In the special case of two-pixel images the effects of noise and the model transform ($S$) can be illustrated in the two-dimensional space of brightness and contrast. 



\begin{figure}[b!]
\begin{center}
   \includegraphics[width=0.98\textwidth]{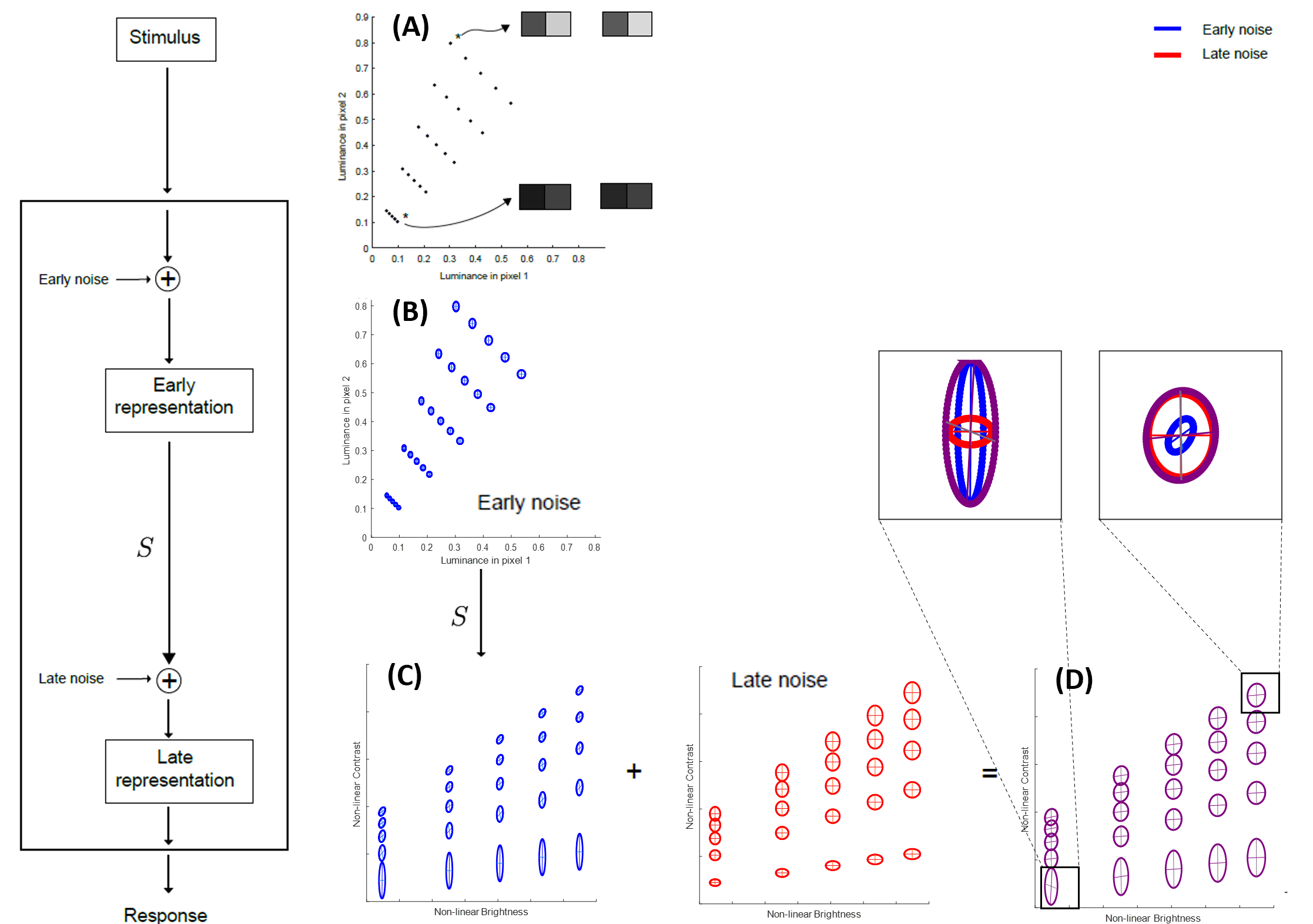}\\ [-0.0cm]
   \caption{\textbf{Modelling framework: propagation of stimulus and inner noise through the system in the absence of external noise.} (A) In this toy example input stimuli consist of two-pixels of varying luminance (axes) shown at their corresponding positions in the 2D coordinate system. 
   (B) Early stimulus representation with added early noise (blue). Ellipses indicate the magnitude of variation resulting from the added noise. (C) The early representation goes through the fixed, deterministic vision model ($S$), which results in non-linear transformations of the output. The x-axis is now non-linear brightness and the y-axis non-linear contrast. Note the different position and orientation of the blue ellipses. Then, late noise is added (red ellipses at the same positions as the stimulus representations). (D) Late representation of stimulus and inner (early + late) noise, which limits discrimination performance. \blue{The  standard deviations and fano factors control the size of the ellipsoids. The specific values were chosen to be illustrative taking into account that in this example the range of luminances is normalized to [0,1]. 
   The interested reader can access and edit the expressions of the different noise sources of this two-pixel model available online in the aforementioned web site.}}
\label{twopixel_basic}
\end{center}
\end{figure}

Figure \ref{twopixel_basic}A shows the representation of two-pixel stimuli varying in their respective luminances in a coordinate system where the x- and y-axis define the luminances of pixels 1 and 2, respectively. Stimulus variations  along the main diagonal reflect variations in mean luminance as both pixels have the same luminance (from low in the lower left to high in the upper right). Variations along the other diagonal (orthogonal to mean luminance) represent variations in contrast. Contrast along the main diagonal is zero and increases towards the upper left and lower right corners. We sampled the two-pixel images at constant steps in luminance. This will help to see the nonlinear deformation of the noisy stimulus representation before and after application of the model transform $S$.  

Figure \ref{twopixel_basic}B shows the noisy representation of the stimulus after adding early noise (\textit{blue} ellipses indicate one standard deviation in each direction). The area of the ellipses increases with increasing stimulus level, because we use level-dependent noise. 
Specifically, we defined the early noise as independent, Poisson-like noise, following \citet{Cotaris1,Cotaris2} and \citet{Dayan01}. 
For this type of noise the horizontal and vertical widths of the blue ellipses increase with luminance, in agreement with the fact that the standard deviation of Poisson noise increases with the signal. 
The main diagonals of the blue ellipses are parallel to each of the axes in Figure \ref{twopixel_basic}B, respectively, because there is no correlation between adjacent sensors---the Poisson-like noise is independent at every sensor response. 
This \textit{early representation} is the input to the model transform $S$. 

Figure \ref{twopixel_basic}C shows the representation \textit{after} the model $S$ has been applied to the early representation.  
The coordinate axes are now labeled brightness and contrast, respectively, because the linear transform $W$ had the effect of rotating the original pixel space by 45 degrees. In the new coordinate system the axes roughly correspond to these perceptual variables. 
After the masking transform $m$, the distance between stimuli is no longer equal. Stimuli of low luminance or low contrast are now further away from each other, and stimuli of higher luminance or contrast are closer. This is consistent with Weber's law \citep{Fairchild13} and with the reduced discriminability of high-contrast stimuli \citep{Nachmias_1974,Legge80,Legge81}. 
The transformation of distances resulting from $S$ depends on the Jacobian of the deterministic transform $S$. 

Figure \ref{twopixel_basic}D shows the representation of the model output after late noise (ellipses in \emph{red}) was added (Fig. \ref{twopixel_basic}C).
Similar to the early noise, the late noise is chosen to be independent and Poisson-like (red ellipses in Fig. \ref{twopixel_basic}). 

The contributions of early (passed through $S$) and late noise combined define the inner noise (ellipses in \emph{purple}).
To illustrate the separate contributions of early and late noise to the late (stimulus+inner noise) representation in Figure~\ref{twopixel_basic}D, and their respective effects on discrimination behavior, the purple ellipses are enlarged for two stimuli (zoom-insets).   
At low brightness and contrast (left inset), variability (inner noise, purple ellipses) in the late representation is dominated by early noise (blue and purple ellipses are very similar).
At high brightness and contrast (right inset), variability in the late representation is fully determined by late noise (red and purple ellipses are almost identical).

Discrimination between two stimuli, $\vect{x}$ and $\vect{x}+\Delta\vect{x}$, is possible when the Euclidean distance in the late representation, $\Delta S = S(\vect{x}+\Delta \vect{x}) - S(\vect{x})$, is larger than the standard deviation of the inner noise in the same direction. 
Thus, discriminability depends not only on the distance between two points in the late representation, but also on the distribution of the inner noise (ellipses in purple). 
These two factors can be taken into account simultaneously by using the multivariate Mahalanobis metric \citep{Maha36}.  
We derive the theoretical Mahalanobis metric for stimuli passed through a vision model ($S$) with known early and late noise magnitudes, 
and compare these with an empirically obtained Mahalanobis metric---the empirical detection and discrimination thresholds---for the same stimuli. 
We can repeat this for several early and late noise magnitudes, and find the values that best account for what is observed experimentally. 
This is the \textbf{first core idea} of our method. 




\subsection{2.4 External noise as a tool to estimate inner noise}

\noindent In addition to the early and late noise sources of the visual system, we consider an extra noise source, the \textbf{\emph{external noise}}, $\vect{n}_\varepsilon$. External noise is under the control of the experimenter and will be added to the stimulus as the tool to gauge the amount of early and late noise.
Adding external noise expands the scenario described in Eq.~\ref{noisy_resp} in the following way:
\vspace{-0.1cm}
\begin{eqnarray}
  \xymatrixcolsep{2pc}
  \xymatrix{
     \vect{x} + \vect{n}_\varepsilon + \vect{n}_e
  \,\,\,\,\,\,\, \ar@/^2.3pc/[r]^{\scalebox{1.0}{$S$}} & 
  \,\,\,\,\,\,\,\,\,\,\,\,\,\,\,\,\,\,\,\,\, \vect{y} = S(\vect{x} + \vect{n}_\varepsilon + \vect{n}_e) + \vect{n}_l
  }
 \label{noisy_resp2} \\
  & \!\!\!\!\!\!\!\!\!\!\!\!\!\!\!\!\!\!\!\!\!\!\!\!\!\!\!\!\!\!\!\!\!\!\!\!\!\!\!\!\!\!\!\!\!\!\!\!\!\!\!\!\!\!\!\!\!\!\!\!\!\!\!\!\!\!\!\!\!\!\!\!\!\!\!\!\!\!\!\!\!\!\!\!\!\!\!\!\!\!\!\!\!\!\!\!\!\!\!\!\!\!\!\!\!\!\!\!\!
  \vect{y} = S(\vect{x}) + \vect{n}_\mathcal{I} \nonumber
\end{eqnarray}
where now the inner noise $\vect{n}_\mathcal{I}$ also includes a contribution of the external noise. This modification can be applied to any model, for instance to our vision model $S$ considered in Eq.~\ref{modular}.



\begin{figure}[t]
\centering
    \includegraphics[width=0.98\textwidth]{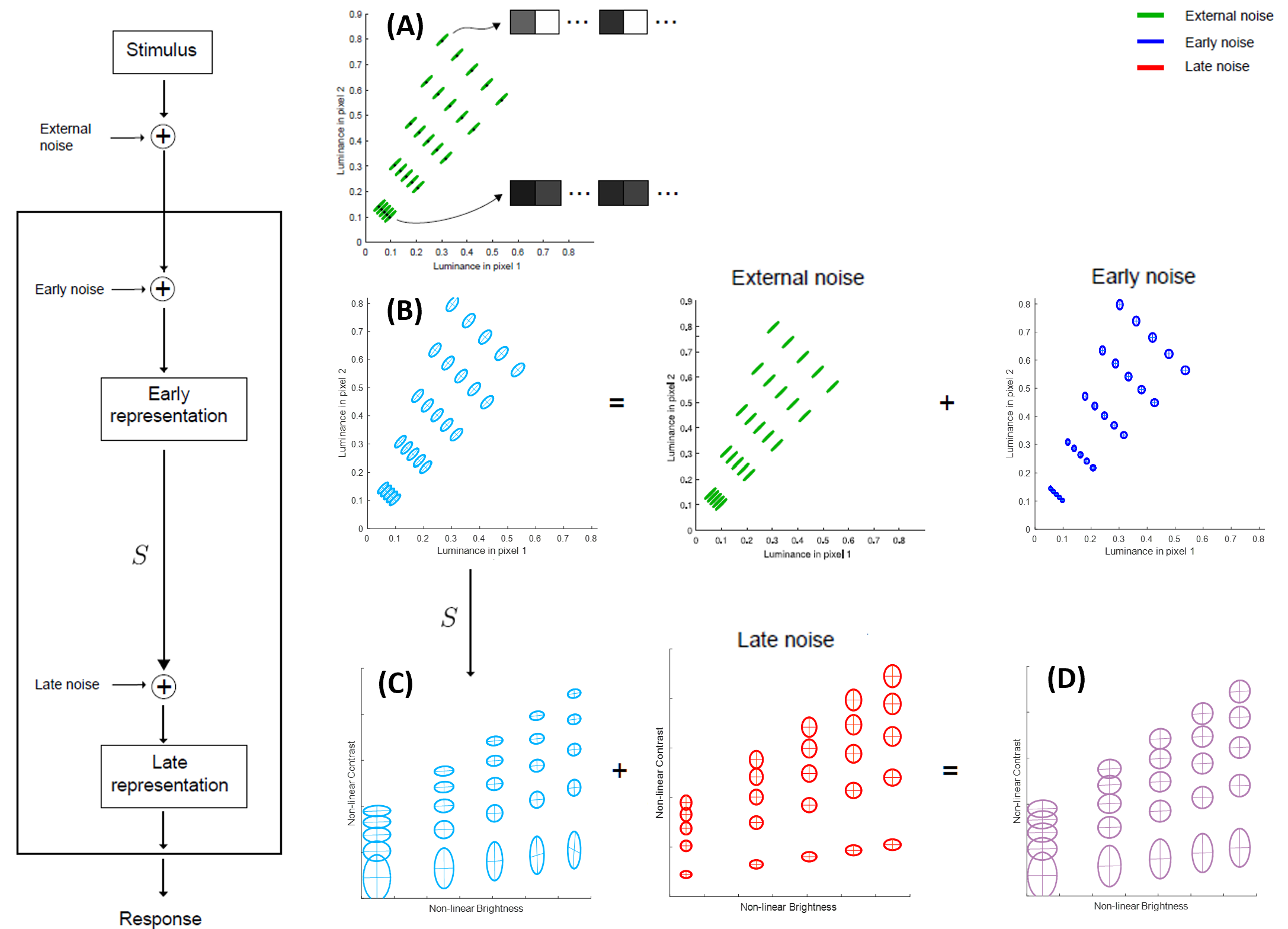}
   \caption{\textbf{Modelling framework: propagation of stimulus and noise through the system in the presence of external noise.} The Figure is organized analogous to Figure 1. (A) External noise is applied to the luminances of each pixel so that the values of the two luminances vary around their mean (green ellipses). 
   \blue{(B) Noise at early representation (cyan ellipses) comes from the superposition of external noise (green ellipses) with early noise (blue ellipses). (C) The early representation goes through the vision model, $S$, as in the previous figure. Then, late noise is added (red ellipses). (D) Late representation of noisy stimuli, with external, early and late contributions.}}
   \label{twopixel_external}
\end{figure}

Figure~\ref{twopixel_external} illustrates the effects of external noise in our modeling framework for our toy example with two-pixel stimuli.
\emph{Green} ellipses indicate the luminance variations in the two-pixel stimuli which depend on the type and amount of external noise applied in a particular experiment. 
Here we use pink noise as external noise, which has a 1/f amplitude spectrum~\citep{Wichmann02}. This type of external noise has more energy (induces larger variation) in the mean luminance direction and less energy (smaller variation) in the contrast direction. 
It has the same covariance matrix (the same energy) for every stimulus across the stimulus space. These two features are evident in the stimulus representation (Figure \ref{twopixel_external}A). Green ellipses are elongated in the main diagonal (more variation in mean luminance), and all ellipses have the same size.\footnote{The statistical properties of the external noise are the calibration features which the experimenter may choose to probe hypotheses about the system under study. For instance, changing the spatial spectrum of the external noise is equivalent to introducing correlations over the different locations (pixels), and hence changing the orientation of the green ellipses.}.

\begin{figure}
\centering
\includegraphics[width=0.5\textwidth]{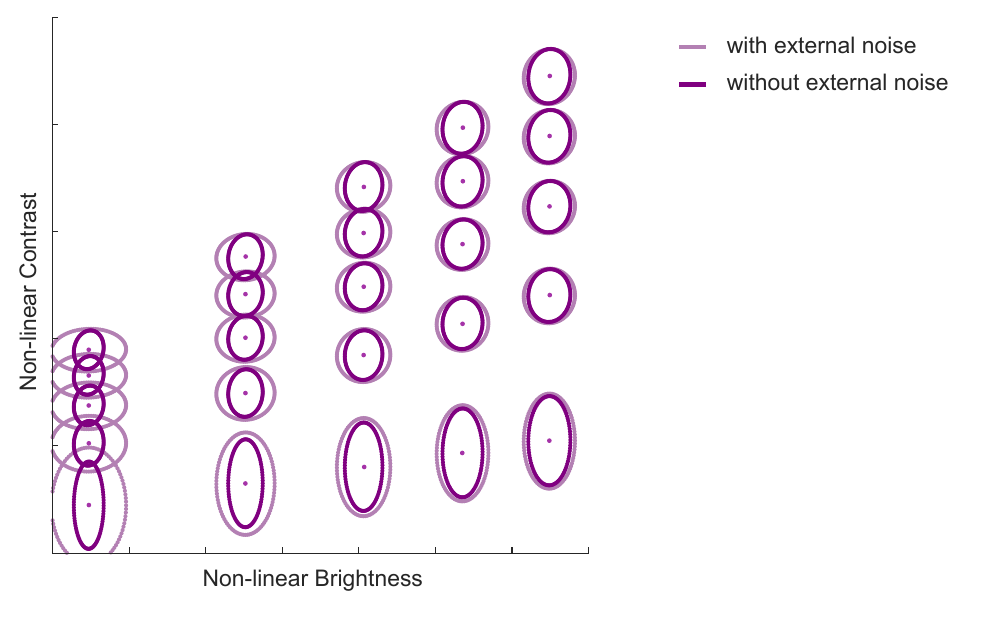}
\caption{\textbf{Late representation with and without external noise.}
Same as in Fig. \ref{twopixel_basic}D and \ref{twopixel_external}D, but visualized together for easy comparison.
Differences imply that controlled variations in the external noise induce different variations of the thresholds for different pedestals and directions.}
\label{twopixel_with_without}
\end{figure}

Figure \ref{twopixel_external}B shows the stimulus representation (\emph{light-blue} ellipses) in the presence of early noise (dark blue). Early noise and late noise are identical to the previous case (Fig. \ref{twopixel_basic}B). Note that this time, however, the external and early noise are summed together and commonly contribute to the noisy early representation. 
Figure~\ref{twopixel_external}C shows how the noisy stimulus representations are transformed by $S$, again leading to a deformation of the ellipses.  
Late noise (red ellipses) is added in the same way as in the previous case resulting in the late representation (\emph{light-purple} in Fig.~\ref{twopixel_external}D).
In this scenario the late representation also includes external noise in addition to inner noise, which is evident from overall larger ellipses than in the scenario without external noise. This is illustrated in Figure~\ref{twopixel_with_without}, which shows the late inner representation of the stimuli in the presence (light violet) and absence of external noise (dark violet). 

The pink noise we used here has higher energy along the mean luminance direction (see above). Hence,  differences in \emph{variability} between late representations with and without external noise are larger along the brightness axis than along the contrast axis (see Fig.~\ref{twopixel_with_without}). Also, this type of external noise has its largest effect on stimuli with low brightness and low contrast (lower left corner of Fig.~\ref{twopixel_with_without}). It is evident from the Figure, that the inner noise ellipses of the two lowest contrast stimuli at the lowest brightness level have substantial overlap in the presence of external noise (light violet), and would hence be less discriminable than without external noise. The two highest contrast noise ellipses at the highest brightness level did not change with respect to their overlap. The change in discriminability as a function of varying types and/or magnitudes of external noise, is the critical piece of information which renders external noise the gauging stick to calibrate the scale of  internal noise.

This is the \textbf{second core idea} in our noise estimation method which we will describe in what follows.
The theory allows us to compute the covariance matrices which represent the noise-induced variability, and their associated ellipses, at the different levels of representation, that is, \textit{before and after the transformation} $S$.  
We can calculate ellipses, i.e. the magnitude, of the three noise components (external, early, and late) at every representation level, regardless of where they actually arise. 
This means that we can \textit{go back and forth} through the system, and can for instance, express the inner noise and its components in the coordinates of the stimulus. This allows us to study the effect of noise on performance not only at the late representation, but also to project the noise back to the stimulus space and predict discrimination performance in units of the stimulus (the psychometric function, see below).

\subsection{2.5 Interim summary to our method's intuition}
\noindent We used a toy example of two-pixel images to illustrate the two core ideas of our method.   
First, discrimination between stimuli depends on two factors, their \textbf{distance} in the late representation and their \textbf{variability}, both taken into account by the Mahalanobis metric.
Two stimuli $\vect{x}_A$ and $\vect{x}_B$ can be discriminated if their deterministic responses fulfill the following criterion: Given one of the deterministic response vectors, say $S(\vect{x}_A)$, the other, $S(\vect{x}_B)$ is far enough away from $S(\vect{x}_A)$ that it cannot be confused with its noisy version $S(\vect{x}_A) + \vect{n}_\mathcal{I}(\vect{x}_A)$. 
Second, using noise propagation theory and metric transformations, we can represent external and early noise at the level of the late representation. Together the different noise sources determine discrimination performance at this level of representation. 

In what follows we present two versions of our approach to estimate early and late noise from psychophysical data.   
Section 3 explains a first approach using threshold-only data. We show how to estimate early and late noise contributions in the presence of external noise. 
Section 4 presents a second, simulation-based non-parametric approach. Here we derive early and late noise contributions as well, but this time we use the full psychometric function and thus do not need external noise. 


\subsection{2.6 Experimental data used for modelling}
\noindent We apply our proposed method to detection and discrimination data with well-controlled external noise reported by~\citet{Wichmann02}. \citet{Wichmann02} compared the detection and discrimination of standard sine gratings with so-called pulse-train stimuli, containing not only energy at the fundamental frequency---as in sine gratings---but equally much at all harmonics (within the limits of the display system). Both detection and discrimination experiments were repeated with and without added external pink (1/f) noise.
\blue{Experiments were conducted as two-alternative forced-choice at a presentation duration of 79 msec within a rectangular temporal window. All gratings were horizontally oriented within a spatial Hanning window nominally subtending 3.8 deg at the observers' eyes. Data were collected using the method of constant stimulus with 50 trials per block with 500 to 600 trials in total per psychometric function.}

We chose to model the data from \citet{Wichmann02} because in their paper the authors argue that the pattern of results they obtained are consistent with the notion that low-contrast detection is limited by an early inner noise source but high-contrast discrimination is limited by late inner noise. They thus experimentally addressed the very same question we now believe to be able to address theoretically and computationally---hence we felt it was appropriate to use their data for our analysis.

\blue{All our simulations below use exactly the same stimuli as in the original experiments because we
are running the same code to generate the images and the pink noise. This code is available online in the aforementioned web site.}

\section{3. Noise estimation I: Using only threshold data}
\subsection{3.1 Theory: perceptual distance in terms of thresholds and noise}

\blue{The key of our proposal is to relate the psychophysical thresholds for detection and discrimination with or without external noise to a perceptual metric derived from a noisy nonlinear model.
In a discrimination experiment, where a target stimulus $\vect{x}$ is varied (``distorted'') in one direction of image space $\Delta\vect{x}$, experimental thresholds in that direction $|\Delta\vect{x}|_{\tau}$ define an \emph{empirical perceptual distance}:} 
\begin{equation}
      D_{\textrm{emp}} = \frac{1}{|\Delta\vect{x}|_{\tau}}
      \label{D_emp}
\end{equation}
\blue{As a consequence, a high threshold---a large $|\Delta\vect{x}|_{\tau}$---indicates a small perceptual distance in that direction of image space: the stimulus has to change a lot in this direction before an observer is able to discriminate the original from the distorted stimulus.}

\blue{However, in the previous section, we suggested that the departure in the inner domain is not the only determinant of discriminability, but also the inner noise. Therefore, our proposal for the \emph{theoretical perceptual distance} use the Mahalanobis metric as it simultaneously takes both factors into account~\citep{Maha36}: the departure in the response, $\Delta S$, weighted by the covariance of the inner noise:}
\begin{equation}
       D_{\textrm{th}}^2 = \Delta S^\top \cdot \left( \Sigma_{\mathcal{I}}  \right)^{-1} \cdot \Delta S
       \label{maha_in_response}
\end{equation}
\blue{where $\Delta S = S(\vect{x}+\Delta \vect{x}) - S(\vect{x})$, and $\Sigma_{\mathcal{I}}$ is the covariance matrix of $\vect{n}_\mathcal{I}$.} 

In order to express this proposed theoretical distance in terms of (1) the stimulus, and (2) the different noise sources---\emph{external}, \emph{early} and \emph{late} noise---we invoke two known results: First, the change of the perceptual metric matrix under deterministic transforms~\citep{Malo06a},
and, second, the change of the covariance of the noise under deterministic transforms~\citep{Ahumada87}. Both of these two results use the Taylor approximation of the nonlinear behavior of the system, $S$. This assumption is correct if the nonlinearity of the system is moderate, or if the noise is smallish relative to the signal (low-noise limit).

Applying the work of \citet{Malo06a} and \citet{Ahumada87} to equation~\ref{maha_in_response}, we see how the different noise sources contribute to the theoretical perceptual distance induced by a variation in the stimulus $\Delta \vect{x}$ 
(see Appendix A for derivation):
\begin{equation}
      \hspace{-0.5cm}
      D_{\textrm{th}}^2 = \Delta\vect{x}^\top \nabla S^\top
      \left(
      \underbrace{k^2 \, \Sigma_l}_{\textit{late noise}} +
      \underbrace{k^2 \, \nabla S \Sigma_e \nabla S^\top }_{\textit{early noise}} +
      \underbrace{2 k^2 \mathbb{E}[\vect{n}_l \vect{n}_e^\top \nabla S^\top] }_{\textit{corr. late-early}} +
      \overbrace{
      \underbrace{\nabla S \Sigma_\varepsilon \nabla S^\top }_{\textit{external noise}} +
      \underbrace{2 k \nabla S \mathbb{E}[\vect{n}_e \vect{n}_\varepsilon^\top] \nabla S^\top }_{\textit{corr. early-external}} +
      \underbrace{2 k \mathbb{E}[\vect{n}_l \vect{n}_\varepsilon^\top \nabla S^\top] }_{\textit{corr. late-external}}
      }^{\textit{external dependent}}
      \right)^{-1} \!\!\!\!\!\!  \nabla S \Delta\vect{x}
      \label{metric_noise1}
\end{equation}

\noindent where $\nabla S$ is the Jacobian of the model at $\vect{x}$ and the matrix $\Sigma_l$ is the covariance of the late noise $\vect{n}_l$ at the point $S(\vect{x})$. The term that depends on $\Sigma_{e}$ describes the early noise at $\vect{x}$ propagated up to the late representation. 
In addition, we introduced an arbitrary scale factor, $k$, on all the observer's noise sources (early and late) to indicate the critical role the \emph{external} noise with covariance $\Sigma_\varepsilon$ plays to obtain the scale of the uncertainties---we will return to this issue below. 
The terms with the expected value, $\mathbb{E}[\cdot]$, are all zero in case of independent noise sources. However, if noise sources are dependent the terms with $\mathbb{E}[\cdot]$ are non-zero. In general they are not zero: for instance, in the conventional Poisson choice we have made here, the different noises depend on the signal and thus the cross correlation matrices do not vanish. The cross correlation terms imply that the covariance of the sum of the contributions to the inner noise is not simply the sum of individual covariance matrices. The proper \emph{combination} of the noise sources has to be calculated.


\subsection{3.2 Generality of Eq.~\ref{metric_noise1} and derivation of classical expressions as special cases}

\noindent Equation~\ref{metric_noise1} is general insofar as it makes no assumptions about the nature of the noise sources or their (in)dependence. Moreover, it can be applied to any model $S$---thus equation~\ref{metric_noise1} holds for more complex (spatial vision) models than the one we consider here and introduced in section 2.2. It also holds if one were to explore different noise models than the Poisson and pink noises we consider here (and in the particular cases further developed in Appendix A). The validity of Eq.~\ref{metric_noise1} is only tied to the Taylor expansions in~\citet{Malo06a} and \citet{Ahumada87}, and these expansions are reasonable in case of low noise and moderately non-linear systems.

From our general expression in Eq.~\ref{metric_noise1} one may deduce classic special cases if one considers the appropriate restrictions in the system and in the noise sources. One such special case is that of \citet{Burgess88}: A linear shift-invariant systems \emph{without} early noise, and stationary and signal-independent \emph{external} and \emph{late} noises. In our terminology, in \citet{Burgess88} we have zero early noise, $\vect{n}_e = 0$ and the late noise $\vect{n}_l$ is not Poisson but is constant, e.g. Gaussian with constant variance independent of the input $\vect{x}$ to the system $S$. This considerably simplifies equation~\ref{metric_noise1} as all terms with $\vect{n}_e$ vanish as well as all those with $\mathbb{E}[\cdot]$.

Moreover, we can derive the special case for a linear shift-invariant system in the Fourier domain. For such a system, its Jacobian can be written as $\nabla S = F^\top \cdot \lambda \cdot F = \nabla S^\top$, where $F$ is a matrix with the Fourier basis functions (in rows), and $\lambda$ is a diagonal matrix with the weights that are applied to each spatial frequency (the filter that represents the system). The covariance of stationary noises which do not depend on the signal can be formulated in the Fourier domain as well: $\Sigma_{\varepsilon} = F^\top \cdot N_\varepsilon \cdot F$ and $\Sigma_l = F^\top \cdot k^2 N_l \cdot F$, where $N_\varepsilon$ is a diagonal matrix with the energy spectrum of the external noise, and $k^2 N_l$ is the corresponding energy spectrum of the late noise (where we kept the scaling factor $k$ to denote its amplitude).

Introducing the above special cases in Eq.~\ref{metric_noise1}, taking $\Sigma_e = 0$ and $\vect{n}_e = 0$, using the orthogonality of the Fourier matrix, and considering the values (in the Fourier domain) in the diagonals of $\lambda$, $N_l$ and $N_\varepsilon$, we finally have:
\begin{equation}
    D_{\textrm{th}}^2 =
    \Delta\vect{x}^\top \cdot F^\top \cdot \left( \frac{\lambda^2}{k^2 \, N_l + \lambda^2 \cdot N_\varepsilon} \right)\cdot F \cdot \Delta\vect{x}
    \label{classic}
\end{equation}
which is equivalent to equation~(1a) in \citet{Burgess88}.


\subsection{3.3 The special role of the external noise} 
Another interesting feature of Eq.~\ref{metric_noise1} is that it highlights the critical role of the external noise as a calibration parameter (or \emph{reference}) to determine the scale of the other noise sources in the system (the subject of our study).
In the absence of such a reference, the other noise sources could be arbitrarily scaled with no impact on the correlation between theory and experiment.

Note that if no external noise were used in the experiments, the term which does not depend on $k$ disappears. As a result, the theoretical perceptual distance reduces to:

\begin{equation}
      D_{\textrm{th}}^2 = k^{-2} \left(\Delta\vect{x}^\top \nabla S^\top
      \left(
       \Sigma_l +
       \nabla S \Sigma_e \nabla S^\top  +
       2 \mathbb{E}[\vect{n}_l \vect{n}_e^\top \nabla S^\top] 
       \right)^{-1} \nabla S \Delta\vect{x} \right)
      \nonumber
\end{equation}
This means that an arbitrary scaling $k$ of the size of the noise sources only leads to a corresponding scaling $k^{-2}$ of the theoretical distance, which has no effect on the correlation between  $D_{\textrm{th}}$ and $D_{\textrm{exp}}$.

As a consequence, by maximizing the correlation with no external noise one could fit the structure of the covariance of the noise sources, but not its absolute scale, which would be of limited interest---external noise effectively anchors the absolute scale of the internal noises (early and late) relative to the \emph{known} external noise.

The sum of terms in Eq.~\ref{metric_noise1} implies that the different noise sources actually play the role of relative references for each other: global scaling of all the noise sources at the same time has no effect on the correlation, but considering a bigger contribution (say of late noise) with no variation of the other sources, would have an impact in reproducing the experiments.

This observation
suggests that the experiments used to determine the noise should use substantial amount of external noise and include early noise in the formulation to ensure the necessary constraints for the scale of the late noise and even noise at the (putative) decision stage.
\blue{This role of the external noise as scale was also apparent in the classical Eq.~\ref{classic}: in absence of external noise, $N_\varepsilon = 0$, the scale of the late noise $k^2$ is arbitrary, as it goes out of the denominator and just scales the distance. The advantage in our Eq.~\ref{metric_noise1} is that it explicitly contains the early noise, so it can also be compared to the external noise.}
This special role is the reason to use accurate threshold measurements in external noise such as those in~\citet{Wichmann02}.


\subsection{3.4 Estimation I}

Our proposal for noise estimation stated in section 2.3 consists of finding the noise parameters that maximize the correlation between $D_{\textrm{exp}}$ (Eq.~\ref{D_emp}) and  $D_{\textrm{th}}$  (Eq.~\ref{metric_noise1}, which depends on the noise).
In principle, given $n$ experimental conditions to measure $n$ thresholds, this optimization reduces to computing $\frac{\delta D_{\textrm{th}}^i}{\delta \theta}$, where $\theta$ are the noise parameters and $D_{\textrm{th}}^i$ is the distance for the $i$-th experimental condition, with $i=1,\ldots,n$~\citep{Martinez18}.

However, the inverse in Eq.~\ref{metric_noise1} poses serious computational problems.
Note that the derivative of the distance depends on the derivative of the metric, and the inverse implies a dependence on $(\Sigma_\mathcal{I})^{-2}$.
Computing the inverses of very large matrices on every iteration of the optimization is not feasible in practice\footnote{Note that dealing with $64\times64$ images and a 3-scales and 4-orientations steerable transform implies working with metric matrices of size $25664\times25664$. Optimization takes about 20-50 iterations and in each iteration we need to compute $n$ of such inverses (one per data point).} unless
strong restrictions in the nature of the noise are assumed\footnote{For instance, if late noise is assumed to be independent of early noise -true in the low-noise limit-, and we have
restricted Poisson (i.e. only depending on the global energy of the response and not on the energy of each coefficient), the covariance of the transformed input noise can be diagonalized, the corresponding orthogonal matrices can be extracted from the inverse and the inverse reduces to inverting a diagonal matrix. 
}.

Therefore, while the parametric expression presented above in Eq.~\ref{metric_noise1} is helpful to understand the problem---for instance to link it to the classical results, or to understand the relevance of the external noise as a scaling factor for the unknowns---in practice the optimization is easier by taking a nonparametric computation of the theoretical distance. 
We consider the difference in response and the noise at the same time by taking the 
expected value of the Euclidean distance 
between the noisy responses, $\vect{y}(\vect{x})$ and $\vect{y} (\vect{x}+\Delta\vect{x}) = \vect{y(\vect{x})} + \Delta\vect{y}$:
\begin{eqnarray}
      D_{\textrm{th}}^2 &=& \mathbb{E}\left[ \, \Delta\vect{y}^\top\cdot\Delta\vect{y}  \,  \right] \nonumber \\
      &=& \mathbb{E}\left[  \,\,\,   \left|\Delta S + \vect{n'}_{\,\mathcal{I}} - \vect{n}_{\mathcal{I}} \right|^2  \,\,   \right]
      \label{distance_non_param}
\end{eqnarray}
where $\Delta S = S(\vect{x}+\Delta\vect{x}) - S(\vect{x})$ is the difference between the deterministic responses,  $\vect{n}_{\mathcal{I}}$ and $\vect{n'}_{\mathcal{I}}$ are different realizations of the inner noise at the points $S(\vect{x})$ and $S(\vect{x}+\Delta\vect{x})$, respectively, and $\mathbb{E}[\cdot]$ refers to expectation.
Eq.~\ref{distance_non_param} implies that, when judging the difference between two stimuli, the observer compares two noisy responses: $S(\vect{x})+\vect{n}_{\,\mathcal{I}}$ and
$S(\vect{x}+\Delta\vect{x}) + \vect{n'}_{\,\mathcal{I}}$.

From a technical point of view, 
the non-parametric Eq.~\ref{distance_non_param} is not restricted by the local linear approximation required in the analytical Eq.~\ref{metric_noise1}. Appendix~B shows that derivatives of  Eq.~\ref{distance_non_param} do not involve matrix inversion, so the optimization of the noise parameters to maximize the correlation between this $D_{th}$ and $D_{emp}$ is feasible.

In our results below, \blue{the pair of stimuli with and without the test $\Delta\vect{x}$ to be detected or discriminated were put through the model
and then we computed the distances between the corresponding responses in the inner representation.} $\mathbb{E}[\cdot]$ is estimated as an average over $5000$ realizations of inner noise. Appendix~C shows that this amount is sufficient; in numerical simulations we show that the non-parametric distance based on the average over the noisy samples (Eq.~\ref{distance_non_param}) is equivalent to the parametric distance in Eq.~\ref{metric_noise1}.

\subsection{3.5 Results I}

\begin{figure}
  \begin{center}
\includegraphics[width=0.85\textwidth]{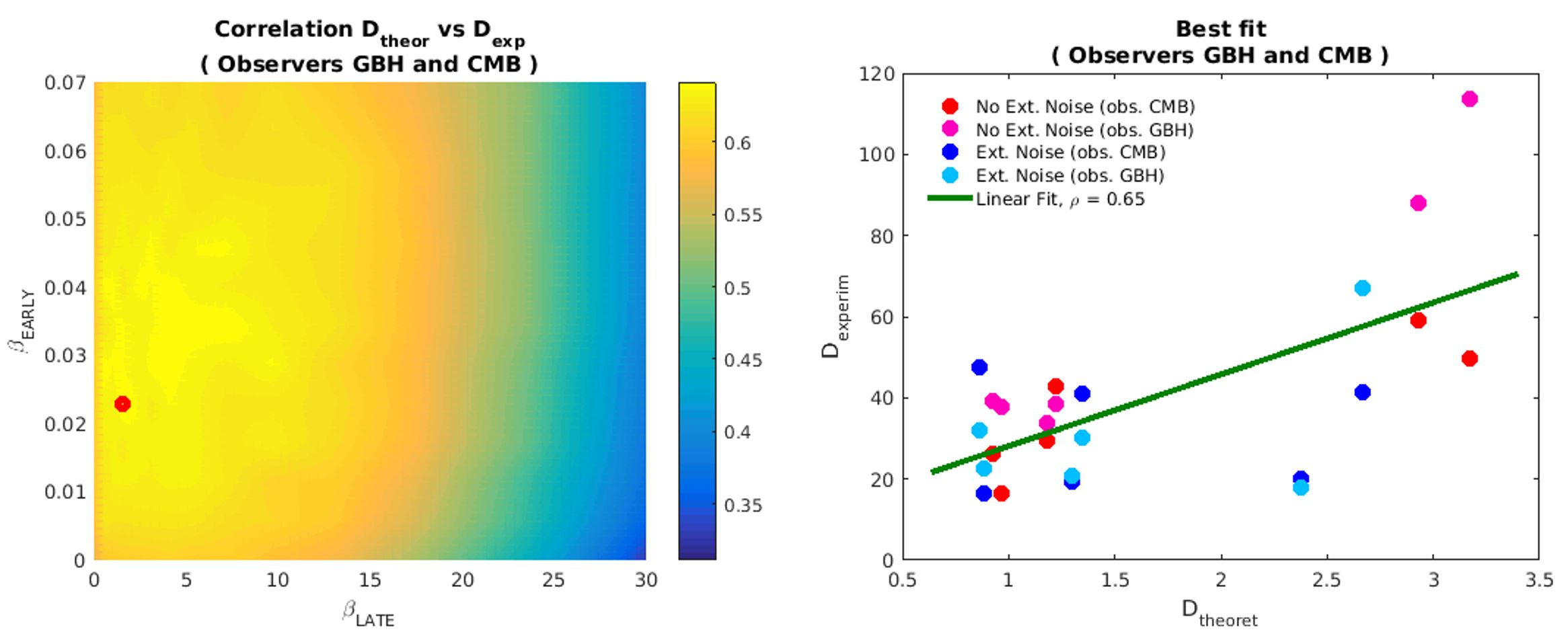}\\   
        [-0.0cm]
        \caption{\small{\textbf{Early and late noise parameters from 
        threshold data.} (Left) Correlation between the theoretical distance (based on the noise) and the experimental distance (based on the thresholds) for various late (x-axis) and early (y-axis) noise Fano factors. The optimal parameters (red dot) that maximize the correlation were $\beta_e = 0.023$ and  $\beta_l = 1.52$. 
        (Right) Scatterplot for the best (maximum) correlation, \blue{for the data of both observers and conditions (with and without external noise) considered together}.
        In this optimal case the Pearson correlation is 0.65. 
        The correlation drops to 0.64 and 0.61 if one neglects either early noise or late noise respectively.
        \blue{As $\beta_l$ is more important to explain the data (has bigger impact in the correlation), $\beta_e$ is less constrained and hence it is more uncertain.}
        }}
        \label{main_result1}
        \end{center}
  \end{figure}

Using the data and model considered above and the computationally convenient noise-dependent distance described in Eq.~\ref{distance_non_param},
we looked for the noise parameters that maximize the correlation between theory and experiment.
The corresponding covariances of the noise are given in Appendix~A. These covariances were used to generate noisy inputs and responses in the non-parametric computation of the distances. 

Fig.~\ref{main_result1} shows the correlations obtained when the noises are assumed to be purely Poisson (with no Gaussian component, and hence $\alpha_e = \alpha_l = 0$). 
The blue-to-yellow color scale represents low-to-high correlations between theory and experiment.
We used all the experimental detection and discrimination data using pulse-trains and sinusoids of different frequencies in external pink noise and with no external noise. 
We fitted the results of both observers at the same time, and the resulting noise parameters were $\beta_e = 0.023$ and $\beta_l = 1.52$.
The figure on the right shows the best (maximum correlation) scatter plot.



%

Fig.~\ref{main_result1} also illustrates the basic problem found in the determination of the noise parameters from the thresholds: the uncertainty of the optimum is large because the correlation surface is very flat.
This problem is also found even if a different set of noise parameters is considered for each individual observer \blue{(see separated results per observer in Appendix D)}. 
When compared to the results in the next section, the use of restricted information (just thresholds as opposed to all the information in the psychometric functions) implies looser constraints and hence bigger uncertainties in the result.

\subsection{3.6 Interim conclusions}

The proposed expression for the theoretical distance, Eq.~\ref{metric_noise1}, has two interesting consequences:
\begin{itemize}
     \item It generalizes the classical result in~\citep{Burgess88} for nonlinear models with sources of noise at different depths (not only signal-independent late noise). 
     \item It points out the special role of the noise at the input.
         Some noise at the input (either \emph{external noise} or \emph{early noise}) is necessary to find the scale of the \emph{late noise}.
\end{itemize}

\section{4. Noise estimation II: Using full psychometric functions}

In this section we follow the \cite{Green_1960} and chapter 5 of \cite{Wichmann_1999}
on the importance of full psychometric functions and not only thresholds for modelling behaviour. In particular, we show that if the full psychometric function is used, there is no need to use external noise in order to scale the size of the early and late noises.

\begin{figure}[h!]
\begin{center}
\includegraphics[width=0.69\textwidth]{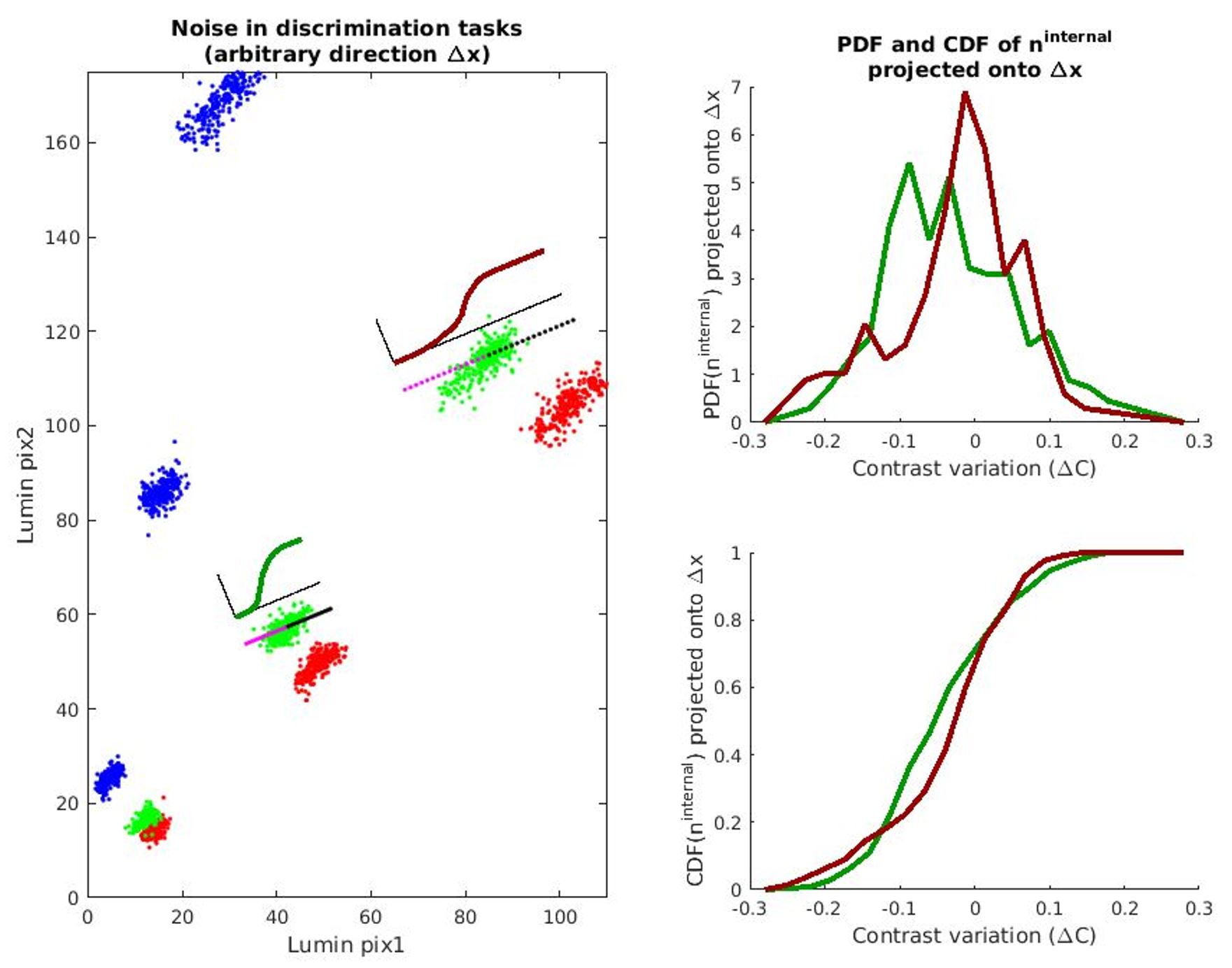}\\         
        [-0.0cm]
        \caption{\small{\textbf{Psychometric functions as a function of the inner noise projected back into the stimulus space} 
        (Left) Two-pixel example (as in Figures~\ref{twopixel_basic}A and ~\ref{twopixel_external}A) showing samples of noisy responses of nine stimuli of different luminance (along the diagonal direction) 
        and different contrast (along the perpendicular-to-the-diagonal direction).
        For two of them we represent departures from the average $\vect{x}$ in a specific direction $\Delta \vect{x}$ with black and pink lines.
        The sigmoids in green and brown represent the CDFs of the noise projected onto the direction of variation of the signal.
        (Right) Probability density functions (PDFs, top-right) and cummulative density functions (CDF, bottom-right) for the two considered clusters.
        Note how the CDF in brown squeezes when expressing the abscisa in contrast units (by dividing the variation by the average luminance).}}
        \label{cdfs_signal_input}
        \end{center}
  \end{figure}

Specifically, we use the definition of the psychometric function in terms of the density of the inner noise given in~\citet{Solomon13}: the probability of correct detection of 
a variation $\Delta \vect{x}$ over the stimulus $\vect{x}$ is given by the cumulative density function of the inner noise in the direction of variation of the stimulus, 
\begin{equation}
      P_{\textrm{correct}} = \frac{1}{2} +\int_{0}^{|\Delta \vect{x}|} p(\vect{u}^\top \cdot \vect{n}^x_\mathcal{I}) \,\,\,\, \vect{u}^\top \cdot d\vect{n}^x_\mathcal{I}
      \label{psychometric}
\end{equation}
where $p(\vect{u}^\top \cdot \vect{n}^x_\mathcal{I})$ is the probability density function (PDF) of the noise in the late representation expressed in the stimulus space and projected in the direction of variation of the signal, $\vect{u} = \frac{\Delta \vect{x}}{|\Delta \vect{x}|}$.
Figure~\ref{cdfs_signal_input} illustrates this concept using the two-pixel example described in section 2.
In this context, the departures from a certain background or pedestal $\vect{x}$ in certain direction $\Delta \vect{x}$ will be visible when the size of the departure can be discriminated over the noise in that specific direction, and this is related to the cumulative density function (CDF) of the noise in that direction (i.e. Eq.~\ref{psychometric}). 

\subsection{4.1 Estimation II}

Given a model that includes noise (e.g. Eq.~\ref{modular}) and discrimination data for increments $\Delta \vect{x}$ over certain pedestals $\vect{x}$
(as the data by~\citet{Wichmann02}), 
we can estimate the noise parameters from the best reproduction of the psychometric functions.
Specifically, the model is used to generate noisy responses for the stimuli and certain noise parameters. Then, we use our noise propagation method to express the noisy responses back in the stimulus domain.
The cumulative histograms of the noisy samples in the direction of the signal variation are the prediction of the experimental psychometric functions for the considered noise parameters.
We then calculate the root-mean square error (RMSE) as a measure of discrepancy between experimental and predicted psychometric functions and use an optimization routine to find the noise parameters that minimize the RMSE. 

Note that at a certain stimulus in Fig.~\ref{cdfs_signal_input} a smaller (bigger) amount of noise or a different orientation of the multivariate PDF would lead to more steeper (shallower) predictions for the corresponding psychometric function. 
In this case (as opposed to Section~3) there is no need of external noise for scaling. Here the variation of the parameters of the 
inner noise directly lead to properly scaled predictions. 


\subsection{4.2 Results II}

\begin{figure}[b!]
  \begin{center}
\includegraphics[width=1\textwidth]{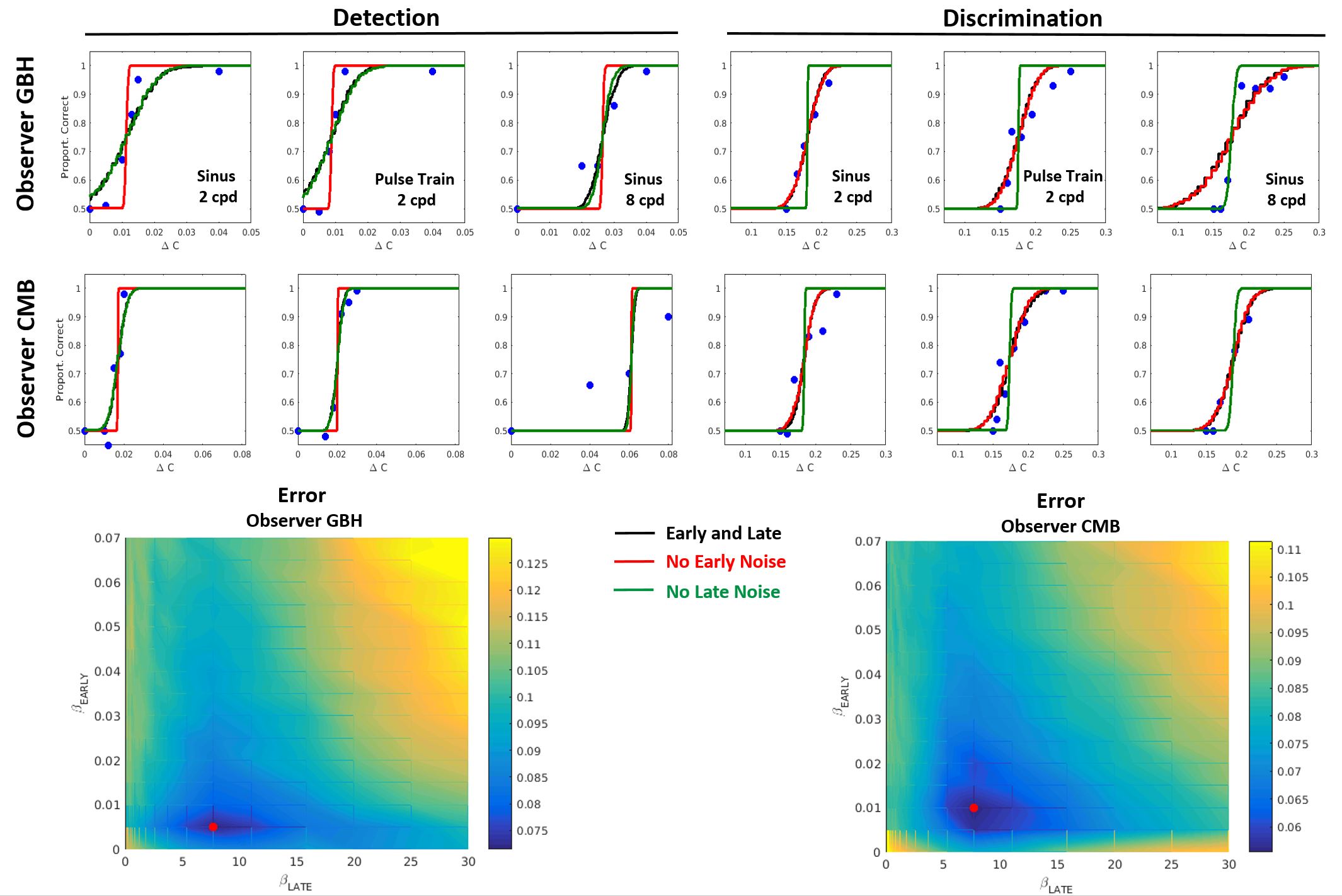}\\[-0.0cm]
    \caption{\small{\textbf{Early and late noise parameters using full psychometric functions.}
        The color surfaces show the average RMSE error between experimental and predicted psychometric functions, for varying noise parameters for the late (x-axis) and early (y-axis) noises. 
        In each case the optimum is given by the minimum in the error surface (marked in red). Please not that the optimum here is a low RMSE error (dark blue) whereas in the previous section the optimum was a maximal correlation (light yellow).
        For the observers GBH and CMB the optima, highligted in red, are located at  
        ($\beta_e = 0.005$, $\beta_l = 7.69$) and ($\beta_e = 0.01$, $\beta_l = 7.69$) respectively.
        The plots at the top represent the experimental data of the psychometric functions (blue dots) and different optimal predictions considering: both early and late Poisson noise (in black), only late noise (in red), and only early noise (in green).}}
        \label{main_result2}
        \end{center}
  \end{figure}

The color surfaces in Figure~\ref{main_result2} show the errors in predicting the psychometric functions for a range of early and late noise parameters obtained with the above procedure for the psychometric function data of the two observers in~\citep{Wichmann02}. We only used and required experimental data without external noise.
The blue-to-yellow color scale represents low-to-high prediction error. 
The optimal noise parameters (minimum error) 
were $\beta_e = 0.005$ and $\beta_l = 7.69$ for observer GBH, and $\beta_e = 0.01$ and $\beta_l = 7.69$ for observer CMB.
The experimental data are shown as blue dots, 
and the continuous red and green curves represent different predictions from the model (Eq.~\ref{psychometric}).
The curves in black correspond to the optimal predictions for each observer using both early and late noise. 
The curves in red and green come from disregarding early noise and late noise respectively. Models with only early or only late noise clearly lead to worse predictions: Both early and late noise is required to account for the experimental data.

Early noise seems particularly relevant in detection, where contrast is small and hence (signal-dependent) late noise is negligible. In detection, neglecting the early noise leads to unrealistic (too steep) psychometric functions (see red curves in Fig. \ref{main_result2}). These predictions (according to the error surfaces) are not improved even if substantially bigger late noise is considered.
On the contrary, early noise seems irrelevant to explain discrimination. In discrimination the neural responses are already high and hence the signal dependent contribution of late noise is much higher than the early noise transformed into the late representation. Note that early noise is compressed by the saturating nonlinearity at high amplitudes in the late representation. 

Late noise seems particularly relevant in discrimination. If late noise is neglected in discrimination the predictions are unrealistic (too steep, see green curves in Fig. \ref{main_result2}).
These predictions (according to the error surfaces) are not improved even if substantially bigger early noise is considered. 
On the contrary, late noise seems irrelevant in detection: note that the no-late-noise green curve matches the black curve in detection.
Explicit examples of the role of early and late noise in the PDF of the inner noise are given in Appendix E.  

Note, however, that the above result and interpretation does not hold in general for all spatial vision models: It is specific for the type of Gaussian-Poisson noise sources assumed in our model as well as for the specific, simple model \emph{S} we use---this is not (yet) the answer to the long-standing question regarding the role of level-(in)dependent late noise in models of spatial vision \citep[see e.g.][]{Wichmann_1999,Kontsevich_2002,Georgeson_2006,Wichmann17}, the issue we return to in the paragraph \emph{Ambiguity between transducer and input-dependent noise} in Section~5.2.

\blue{On the other hand, the full model (black line) does not achieve a perfect fit of the blue dots. This may be due to the limitations of the model $S$ assumed for the illustration: note that the masking
nonlinearity is just a fixed saturation, which is a gross oversimplification. This may certainly limit the maximum performance of the model. More accurate models
including input-dependent saturations, e.g. divisive normalization~\citep{Wichmann17,Martinez18}, could lead to better fits. However, note that the procedure to estimate the noise
contributions propose here is completely general, and also applicable to more accurate models.}

\section{5. Discussion}

We first address the reliability of the proposed methods 
and then we discuss the connections of our psychophysical methods 
with other physiological models and previous psychophysical methods for noise estimation.
Finally, we discuss the implications of our noise estimates on other experimental methods and on information-theoretic approaches to study vision.  
We conclude the work with an overview of our methods and findings.

\subsection{5.1 Reliability of the proposed methods}

In optimization, the accuracy of the estimation is given by the steepness of the goal function 
at the optimum~\citep{Press2007}. In our case, accuracy is substantially higher for the full-psychometric method. Note the clear minima in Fig.~\ref{main_result2} as opposed to the flat plateaus obtained in the threshold-only-method in Figs.~\ref{main_result1} and ~\ref{main_result1_separated}.
See for example the size of the region which leads to a change in the first significant figure of the goal-variable (either correlation or error): the uncertainty regions of the parameters defined in this way~\citep{Press2007} are small in case of using the full psychometric functions and huge in the threshold-only case.
Moreover, the consistency between the results of the two observers is higher in the full-psychometric method with regard to the threshold-only method. In fact, the results of the threshold-only method can be considered compatible only because of their small accuracy (and hence high uncertainty).

\blue{In summary, for the dataset and model explored here, we find improved accuracy and consistency from the full psychometric method over the threshold-only method. However, the data by~\citep{Wichmann02} only contain a single level of external noise, and we can thus not rule out that experiments measuring thresholds only, but at various different noise levels, c.f.~\citep{Legge87,Pelli99,Goris08}, our threshold-only method could also lead to more accurate estimates of the internal noise.}

\subsection{5.2 Related work}

\noindent \textbf{Psychophysical versus physiological estimates of early noise.}
Our psychophysical estimates of the noise should, ideally, have physiological correlates. One might think (naively) that our psychophysical estimate of the early noise is directly related to the noise in the electrical response of L,M,S retinal cones. Similarly, one may think that our late noise could be related to the noise at the cortex.
However, direct comparison is not that simple. 
In fact, following the literature that relates basic psychophysical behavior (such as the Contrast Sensitivity Functions or center-surround sensors) from noise removal goals~\citep{Atick92,Li22}, it makes sense that the psychophysically estimated early noise has substantially less variance than the noise actually happening at the retina. 
Appendix~F shows a comparison between our psychophysical estimates and a reasonable physiological estimation of the retinal noise using an accurate model of the retina, the ISETBio, first used in~\citep{Cotaris1,Cotaris2}). Simulated noise on top of illustrative images in Appendix~F shows reasonable dependence on luminance, contrast and frequency of early and late components of our psychophysical noise and its low variance compared to physiological estimates. This makes sense as the psychophysical noise should be barely visible.
\blue{The exact values of the estimated early and late noises should be interpreted with caution, as they not only depend on the data we used but also on the assumed form of the noises as well as our assumed model. What is important, however, is that the estimates returned by our model are in the right ballpark for a simple but reasonable model and simple but reasonable noise assumptions.}

\noindent \textbf{Ambiguity between transducer and input-dependent noise.}
Our work is focused on the separate determination of the \emph{early} and \emph{late} components of the \emph{inner} noise, so it does not directly address the long-standing question of the ambiguity between the transducer function and the input-dependent variance of the inner noise~\citep{Wichmann_1999, Kontsevich_2002,Georgeson_2006,Garcia09,Kingdom16}.
However, the proposed formulation and nonparametric methods have some connection with that interesting discussion. 
 
First,~\cite{Solomon13} factorize the shape parameter of the psychometric functions into the product of an internal noise component and a transducer component. According to this product, data may be fitted in different (equivalent) ways. 
Our expression, Eq.~\ref{metric_noise1} (or more clearly in the less detailed Eqs.~\ref{inner_in_x} and~\ref{maha2_in_response_AppendixA}), that consider the product of the covariance of the noise and the Jacobian of the transducer are an alternative to the product in~\cite{Solomon13} to describe such ambiguity. 
For instance, in Eqs.~\ref{inner_in_x} and~\ref{maha2_in_response_AppendixA} it is obvious that a variation in the covariance of the inner noise can be compensated by a corresponding variation in the Jacobian, therefore one has an ambiguity. However, Eq.~\ref{metric_noise1} explicitly includes the contributions of the 
early and external noises (not considered by~\cite{Solomon13}) 
and this (1) clarifies that different models $S'$ affect the estimates of the late noise, but not of the early noise, and (2) suggests eventual ways to break the ambiguity.
\blue{On the one hand, consider first a model $S'$ with a linearly scaled sensitivity so that $\nabla S' = \alpha \nabla S$. In that case, Eq.~\ref{metric_noise1} says that the distance $D$ stays the same if the late noise is scaled accordingly, i.e. if $\vect{n'_l} = \alpha \, \vect{n_l}$ and $\Sigma'_l = \alpha^2 \Sigma_l$. 
In the general case with a more complicated $S'$, $\alpha$ will not be constant, but input-dependent. This local/adaptive scaling will affect the late noise in an equivalent adaptive way. However, the (same) early and external noises will be automatically scaled when transferred to the inner representation by $S'$. As a result, only the late noise is affected to keep the relative scale of all the noise sources in the inner domain.}
On the other hand, note that the distances stay the same (equivalent reproduction of the data) only if the Jacobian co-varies with the late noise while the early noise is fixed, which is a rather specific situation.
However, we have not pursued this possibility given the low accuracy of the results obtained using this \emph{threshold-only} method (flat plateaus in Figs.~\ref{main_result1} and~\ref{main_result1_separated}) even for a fixed nonlinearity. 
Similarly to the discussion between~\cite{Kontsevich_2002} and~\cite{Georgeson_2006} accuracy and significance of the results are critical, and hence, it is better not to address this issue with methods that intrinsically have low accuracy, as shown here for the threshold-only method.

In this regard, our non-parametric method that uses the data from all the psychometric functions could be more appropriate for two reasons: one is its better accuracy (illustrated in Fig.~5), and the other is that being non-parametric is not attached to the small-increment (or local linear) assumptions behind Eq.~6, which was one of the factors for the ambiguity pointed out for simple univariate models in~\cite{Kontsevich_2002} (see also the related discussion in chapter 5 by \cite{Wichmann_1999}).

\noindent \textbf{Differences with other psychophysical estimations of the noise.} The Maximum Likelihood Difference Scaling (MLDS)~\citep{Maloney03} is a well founded method to derive nonlinear responses and the size of the inner noise from simple comparisons of suprathreshold distortions. While the default method was designed to estimate signal-independent noise, it can be expanded to incorporate signal-dependent noise~\citep{Kingdom10}. 
Actually, the comparison between the nonlinearities found through MLDS (e.g.~\cite{Mullen22}) and the more traditional integration of incremental thresholds~\citep{Watson97} has been suggested as a way to disentangle 
the nonlinearity and the variability of the inner noise~\citep{Kingdom16}. 
In contrast discrimination tasks similar to the ones considered here MLDS has been found to give substantially higher values for the variance of the noise than simple univariate estimations obtained from the integral of the incremental thresholds~\citep{Mullen22}. 
More importantly, the original and the different variants of MLDS focus on the noise in the response or decision domain, i.e. by construction it does not isolate the relative contributions of the early and late sources of noise.
That is a fundamental difference with the methods proposed here, which explicitly take into account sources of noise at different parts of the model, which is not easy in MLDS. 


The psychometric functions in ~\citet{Solomon13} and \citet{Baldwin16} could be used together with an image computable model in non-parametric ways as we did in section 4.1. 
However, similarly to MLDS, these methods usually subsume all the noise sources in a single one at the inner representation, e.g. what the authors in~\citet{Baldwin16} call equivalent noise, following~\citet{Pelli99}. 
In contrast, here we express metrics and covariance matrices at any depths in the same representation, and this is a way to generalize the models used in the other approaches so that they include uncertainty at different depths along the model.


\noindent \textbf{Our methods can deal with noise at the decision stage.} 
Our Eqs.~\ref{metric_noise1},~\ref{distance_non_param}, and~\ref{psychometric} allow to obtain the contributions to the multivariate inner noise $\vect{n}_\mathcal{I}$. However, previous literature has also considered the univariate uncertainty added to the variable that determines the decision~\citep{Pelli85,Neri10}. 
Appendix~G shows how the proposed methods can be extended to estimate this univariate component as well. Extension is easy because this additional source of uncertainty can be easily incorporated to the covariance of the inner noise. 

\noindent \textbf{Our methods can deal with noise correlations.}
An important difference with all the psychophysical methods considered in the previous paragraph is that our formulation is intrinsically \emph{multivariate}.
The methods mentioned in the section above essentially obtain the noise parameters in the direction where the stimulus is experimentally changed. 
In this way, the consideration of multiple pedestals and incremental stimuli leads to a set of (in principle) disconnected univariate estimations of noise for the different experimental conditions.
In contrast, our multivariate formulations (either the analytic one based on thresholds or the parametric one that uses the psychometric functions) 
are thought to be used with image-computable models and hence whatever multivariate expression of the noise can be assumed and be fitted using all the experimental data at the same time.
In our simulations we considered diagonal covariance matrices for $\Sigma_e$ and $\Sigma_l$ just for simplicity in the explanation, however there is no restriction to assume any correlation in the noise between the sensors at the early or late stages.
This is a fundamental advantage of our method with regard to univariate psychophysical estimations given the relevance of the correlations in the noise, at least for neuroscience ~\citep{MorenoBote14,Averbeck06,Goris_2014,Pouget16}.

\noindent \textbf{Maximum differentiation and noise estimates.} The concept of Maximum Differentiation has been proposed as an experimental way to decide between vision models~\citep{Wang08}, or as a way to measure the free parameters of vision models~\citep{Malo15} using the associated perceptual distance.
Originally~\citet{Wang08} proposed a method to generate stimuli maximally/minimally discernible by observers according to an iterative procedure of perceptual distance maximization/minimization. Depending on the model, this approach may be extremely expensive.
However, following the second order approximation of the non-Euclidean distance (as used in Appendix A) the Maximum Differentiation method was simplified and reduced to the estimation of the eigenvectors of the discrimination ellipsoid~\citep{Malo15,Martinez18}.

The theory proposed here allows the estimation of the covariance matrix of the inner noise and hence the design of interesting stimuli in cardinal directions of the noise-dependent metric. Note that according to the transform of the covariance of noise along nonlinear networks~\citep{Ahumada87}, the directions of maximum and minimum discrimination in the spatial domain depend on the Jacobian of the model and the covariance of the noise in the input:
\begin{equation}
       (\Sigma^{x}_{\mathcal{I}})^{-1} = \nabla S^\top \cdot \Sigma^{-1}_{\mathcal{I}} \cdot \nabla S
       \label{inner_in_x}
\end{equation}
Eigenfunctions of this covariance (where $\nabla S$ and $\Sigma^{-1}_{\mathcal{I}}$ are point-dependent) lead to optimal stimuli in the spatial domain (maximally and minimally discriminable) that can be used to check the effect of the noise components and model parameters at different points of the stimulus space (e.g. detection and discrimination).

\noindent \textbf{Noise estimates in information-theoretic approaches.}
The proposed approach that describes discrimination using a noise-dependent Mahalanobis metric is related with methods that describe discrimination using Fisher information~\citep{Abott99,MorenoBote14,Samengo16,Berardino2017,PNAS24}.
In fact, the linear Fisher information matrix~\citep{Pouget16} has the same expression as our Mahalanobis metric in the input domain (our Eq.~\ref{inner_in_x}).
%
For instance in~\citet{PNAS24} the  Fisher information is $F = \frac{\nabla \mu^2}{\sigma^2}$, where $\mu$ is the deterministic transform and $\sigma^2$ is the variance of the inner noise. Which, in matrix form is 
$F = \nabla \mu^T \cdot \Sigma_{\mathcal{I}}^{-1} \cdot \nabla \mu$, i.e. our Eq.~\ref{inner_in_x}.
Therefore, their expression of the discriminable distance and the corresponding expression for the threshold stimuli are the same as what we get from our Eq.~\ref{maha2_in_response_AppendixA}.
First, from our Eq.~\ref{maha2_in_response_AppendixA} the perceptual distance is proportional to the slope of the response and inversely proportional to the square root of the covariance (as in Eq.~10 in~\citep{PNAS24}).
Second, also from our Eq.~\ref{maha2_in_response_AppendixA}, if a just noticeable variation $\Delta \vect{x}_\tau$ is the one that leads to certain threshold value of the perceptual distance $\Delta \vect{x}_\tau^\top \cdot \nabla S^\top \cdot \Sigma_\mathcal{I}^{-1} \cdot \nabla S \cdot \Delta \vect{x}_\tau = \tau^2$, 
all possible just noticeable distortions lie on an ellipsoid described by $F^{-1}$.
In this way, 
the magnitudes of the just noticeable variations, $\Delta \vect{x}_\tau$, are inversely proportional to the square root of the Fisher information matrix (as discussed after Eq.~10 in~\citet{PNAS24}).
Moreover, the expected value of random just noticeable variations would fulfill 
$\mathbb{E}[\Delta \vect{x}_\tau \cdot \Delta \vect{x}_\tau^\top] = \tau^2 \,\, \left( \nabla S^\top \cdot \Sigma_\mathcal{I}^{-1} \cdot \nabla S \right)^{-1}$. In this case, one would be describing the covariance matrix of the uncertainty of the observer at that point, and it is proportional to the inverse of the linear Fisher information matrix, as in~\citep{Seung93,Pouget16}.


The consistency with the Fisher information results is not surprising given the equivalence of our noise-based Mahalanobis metric with the linear Fisher information matrix. 
However, note that the explicit consideration of the \emph{early}, \emph{late} and \emph{external} noises in $\Sigma_{\mathcal{I}}$, in our Eq.~\ref{metric_noise1}, was not considered in the mentioned information-theoretic literature. As a consequence, our contributions here make the Fisher-information results richer.





\blue{On the other hand, our estimations of the noise have an impact in literature that quantifies information loss in the retina-cortex pathway.
In this regard, values that come from psychophysical data, as those presented here, are more relevant to describe the effective bottleneck than physiological estimates that may be compensated by denoising or enhancing mechanisims~\citep{Atick92,Molano14,Li22}.
Noise estimations are critical either to quantify information flow~\citep{Malo20}, functional connectivity~\citep{Li24}, or to derive subjective distortion metrics based on information loss~\citep{Sheikh06}. In all these applications the estimated variance of the early and late noise can be used for a more accurate account
of the mutual information between the retinal image and the inner representation.}
With the proposed method realistic values for the noise could be obtained for any model, and hence
the results in \citep{Malo20} on the information transmitted by spatial vs chromatic mechanisms or by
linear vs nonlinear transforms could be more than rough estimates with reasonable noise assumptions.
Similarly, better noise estimates could lead to an improvement of image quality measures such
as \citep{Sheikh06} based on modeling information transference along the visual pathway~\citep{Kheravdar21}.

\subsection{5.3 Final remarks: early and late noise in psychophysical models of early vision}

In this work we proposed and compared two methods to obtain the \emph{early} and \emph{late} multivariate contributions to the inner noise that limits visual performance in psychophysical models of early vision. 
Both methods are based on measuring detection and discrimination thresholds. The first method is based on the exclusive use of threshold data, while the second uses the full psychometric functions measured around the thresholds.

The first method, based on the Mahalanobis distance, leads to an analytical expression of the perceptual distance, Eq.~\ref{metric_noise1}, that can be compared with  empirical estimates of the distance based on the thresholds. That expression generalizes~\citep{Burgess88} for any nonlinear model with noise sources at different depths (not only late noise), and points out the fact that the use of external noise in the experiments is strictly required to get the scale of the early and late noise sources when one only used the values of the thresholds.
The second method, which consist of a non-parametric fit of the psychometric functions, points out that the early-noise contribution is strictly required to explain detection results while the late-noise contribution is more important in explaining discrimination (see Fig.~\ref{main_result2}). This is because the inner noise is more similar to the early noise for low-contrast stimuli while it is more similar to the late noise for high-contrast stimuli (see Fig.~\ref{simul_noise2}).  
The second method (that uses more experimental data) is substantially more accurate than the first one. In fact, the results of both methods are compatible due to the large uncertainty in the result of the first method.

Explicit representation of the inner noise in the image domain (in luminance units as in Fig.~\ref{noisesPoisson1}) and its comparison with accurate estimations of photoreceptor noise~\citep{Cotaris1,Cotaris2} shows that our results based on psychophysics make more behavioral sense than results based on retinal physiology, which are way too visible. Our results thus agree with previous suggestions on the role of post-retinal processing as ways to reduce retinal noise~\citep{Atick92,Molano14,Li22,Akbarinia23}.

Given the low accuracy of the threshold-only method with the available data, we did not try to solve the debates on disentagling the noise and the nonlinearities of the models \citep{Wichmann_1999,Kontsevich_2002,Georgeson_2006}. However, Eq.~\ref{metric_noise1} does show that the variation of the inner noise cancels the effect of the transducer (there is ambiguity) only for very specific variations of the early and late noises. This suggests that, with more data to constrain the results the different factors (early, late, and model nonlinearity), could be fitted at the same time.

Beyond the simultaneous estimation of the noise at different depths of the vision model (early, late, and even decision level), other fundamental difference of our work with regard to previous psychophysical methods~\citep{Neri10,Solomon13,Kingdom16,Mullen22} is its multivariate nature. Our expressions allow to accommodate correlations between the different sensors at any depth.
This is important in information-theoretic approaches because some correlations may be more harmful than others~\citep{MorenoBote14,Pouget16}, and we saw that the results of our noise-dependent Mahalanobis metric are equivalent to the results obtained from linear Fisher information matrix~\citep{PNAS24,Seung93}. 

Finally, the proposed psychophysical estimation 
may be used to modify the Euclidean assumptions on the inner metric in recent models of spatial vision~\citep{Wichmann17,Martinez18}.
And this can be used for better quantification of the information flow along the visual pathway~\citep{Malo20}, and to improve image quality metrics based on information loss along the visual pathway~\citep{Sheikh06,Kheravdar21}.

\section{Acknowledgements}

This work was supported in part by MICIIN/FEDER/UE, Spain under Grant PID2020-118071GB-I00 and PDC2021-121522-C21, and in part by the BBVA Programa de Fundamentos de la Ciencia.

\bibliographystyle{apalike}
\bibliography{early_late_references}

\renewcommand{\thefigure}{A\arabic{figure}}
\renewcommand{\theHfigure}{A\arabic{figure}}
\renewcommand{\thetable}{A\arabic{table}}
\renewcommand{\theequation}{A\arabic{equation}}
\setcounter{figure}{0}
\setcounter{table}{0}
\setcounter{equation}{0}
\renewcommand{\thesection}{A\arabic{section}}
\setcounter{section}{0}
\section{Appendix A: Derivation of Eq.~\ref{metric_noise1} (general and particular cases)}
\noindent \textbf{General case.} The theoretical distance in Eq.~\ref{maha_in_response} can be written in terms of the stimulus and the different noise sources by using the Taylor approximation of the nonlinear behavior of the system, as already done in \citep{Ahumada87,Malo99,Malo06a,Laparra10a}, as follows:
\begin{equation}
       S(x+\Delta x) \approx S(x) + \nabla S \cdot \Delta x
       \label{Taylor_AppendixA}
\end{equation}
\noindent where $\nabla S$ is the Jacobian of the model at $\vect{x}$. This Taylor approximation is correct in the low-noise limit or if the nonlinearity of the system is moderate.
Therefore, under this approximation, $\Delta S = S(x+\Delta x) - S(x) = \nabla S \cdot \Delta x$, and $\Delta S^\top = \Delta x^\top \cdot \nabla S^\top$, and the distance in Eq. \ref{maha_in_response} may be written as~\citep{Malo06a}:
\begin{equation}
       D_{\textrm{th}}^2 = \Delta x^\top \cdot \nabla S^\top \cdot \left(\Sigma_{\mathcal{I}}\right)^{-1} \cdot \nabla S \cdot \Delta x
       \label{maha2_in_response_AppendixA}
\end{equation}

Now, by definition, the covariance matrix of the \emph{inner} noise $\Sigma_{\mathcal{I}}$ is given by the expected value of the outer product of the noise: 
\begin{equation}
      \Sigma_{\mathcal{I}} = \mathbb{E}\left[\vect{n}_{\mathcal{I}} \cdot \vect{n}_{\mathcal{I}}^\top\right] = 
      \mathbb{E}\left[(S(\vect{x} + \vect{n}_\varepsilon + k \vect{n}_e) + k \vect{n}_l - S(\vect{x})) \cdot ( S(\vect{x} + \vect{n}_\varepsilon + k  \vect{n}_e) + k \vect{n}_l -S(\vect{x}))^\top\right]
\end{equation}
where, as stated in section 3.1, we included an unknown scale factor, $k$, in the early and the late noise. Using the Taylor expansion again, we have:
\begin{eqnarray}
      \Sigma_{\mathcal{I}} & = &  
      \mathbb{E}\left[( \nabla S \cdot (\vect{n}_\varepsilon +k \vect{n}_e) + k \vect{n}_l ) \cdot ( (\vect{n}_\varepsilon + k \vect{n}_e)^\top \cdot \nabla S^\top  + k \vect{n}_l^\top )\right] \nonumber \\ 
      & = & 
      k^2 \mathbb{E}\left[ \vect{n}_l \cdot \vect{n}_l^\top\right] + 
      k^2 \nabla S \cdot \mathbb{E}\left[ \vect{n}_e \cdot \vect{n}_e^\top\right] \cdot \nabla S^\top +
       2 k \mathbb{E}\left[ \vect{n}_l \cdot \vect{n}_e^\top\right] \cdot \nabla S^\top + \\
      & & + 
      \nabla S \cdot \mathbb{E}\left[ \vect{n}_\varepsilon \cdot \vect{n}_\varepsilon^\top\right] \cdot \nabla S^\top + 2 k \nabla S \cdot \mathbb{E}\left[ \vect{n}_\varepsilon \cdot \vect{n}_e^\top \right] \cdot \nabla S^\top
       + 2 k^2 \mathbb{E}\left[ \vect{n}_l \cdot \vect{n}_\varepsilon^\top\right] \cdot \nabla S^\top \nonumber 
\end{eqnarray}
where the expected value of crossed terms does not vanish in general because (1) the early noise may depend on the signal and hence it may depend on the external noise; and (2) the late noise may depend on the signal and hence may depend on the early and the external noise as well.
In the above equation we can identify the terms of in Eq.~\ref{metric_noise1}:
\begin{equation}
      \hspace{-0.5cm}
      \Sigma_{\mathcal{I}} = 
      \underbrace{k^2 \, \Sigma_l}_{\textit{late noise}} +
      \underbrace{k^2 \, \nabla S \Sigma_e \nabla S^\top }_{\textit{early noise}} +
      \underbrace{2 k^2 \mathbb{E}[\vect{n}_l \vect{n}_e^\top] \nabla S^\top }_{\textit{corr. late-early}} +
      \overbrace{
      \underbrace{\nabla S \Sigma_\varepsilon \nabla S^\top }_{\textit{external noise}} +
      \underbrace{2 k \nabla S \mathbb{E}[\vect{n}_e \vect{n}_\varepsilon^\top] \nabla S^\top }_{\textit{corr. early-external}} +
      \underbrace{2 k \mathbb{E}[\vect{n}_l \vect{n}_\varepsilon^\top] \nabla S^\top }_{\textit{corr. late-external}}
      }^{\textit{external dependent}}
      \label{covariance_inner_noise}
\end{equation}
that represent how the covariance of the noise at the input is propagated through the network~\citep{Ahumada87}.
The above equation contains the (general) covariance matrices of the late noise $\Sigma_l$, the early noise, $\Sigma_e$, and the external noise $\Sigma_\varepsilon$, that may include arbitrary dependence with the signal. The next paragraph makes a particular choice for this variation.

\noindent \textbf{Particular case: Gaussian-Poisson noise.} If \emph{early} and \emph{late} noises are signal-dependent Gaussian-Poisson variables, their realizations are:
\begin{eqnarray}
      \label{Gauss_Poisson_AppendixA}
      \vect{n}_e &=&  \left( \alpha_e I + \beta_e\,\mathbb{D}_{|\vect{x}|^{1/2}} \right) \cdot \vect{n}_{G1}\\
      \vect{n}_l &=&  \left( \alpha_l I + \beta_l\,\mathbb{D}_{|S(\vect{x})|^{1/2}} \right) \cdot \vect{n}_{G2} \nonumber
\end{eqnarray}
where $\vect{n}_{G1}$ and $\vect{n}_{G2}$ denote realizations of unit covariance Gaussian noise in the input space and in the inner space (which have different dimension as there may be a different number of photoreceptors and cortical neurons), and $\mathbb{D}_{(\cdot)}$ stands for a diagonal matrix with the elements of $(\cdot)$ in the diagonal. The corresponding covariance matrices of the different noise sources can be written as:
\begin{eqnarray}
       \Sigma_e(\vect{x}) & = & \alpha_e^2 I + \beta_e^2 \, \mathbb{D}_{|\vect{x}|} + 2 \alpha_e \beta_e \, \mathbb{D}_{|\vect{x}|^{1/2}} \label{Gaussian-PoissonCOV}\\
       \Sigma_l(S(\vect{x})) &=& \alpha_l^2 I + \beta_l^2 \, \mathbb{D}_{|S(\vect{x})|} + 2 \alpha_l \beta_l \, \mathbb{D}_{|S(\vect{x})|^{1/2}}\nonumber
\end{eqnarray}
where $\alpha_e^2$ is the variance of the Gaussian component of the early noise and $\beta_e^2$ is the Fano factor of the Poisson component of the early noise, and equivalent concepts for the late noise.

In this work we assume that both the early and the late noise are pure Poisson sources, a usual assumption in neural systems~\citep{Cotaris1,Cotaris2,Dayan01}, and hence our goal is determining the scale factors $\beta_e$ and $\beta_l$. 

\section{Appendix B: Optimization of correlation from Eq.~\ref{distance_non_param}}

%
%
%

Instead of optimizing the noise through the metric using the mathematical expression for the distance shown in Eq.~\ref{metric_noise1}, which requires the derivative of the metric and, due to the inverse, this implies a dependence on $(\Sigma_\mathcal{I})^{-2}$, we suggest the optimization of the noise using a mathematical expression for the distance based on the difference in the noisy responses: 
\begin{eqnarray}
      \vect{y}(\vect{x}) &=& S(\vect{x})+\vect{n}_{\,\mathcal{I}}\\
      \vect{y}(\vect{x}+\Delta\vect{x}) &=& S(\vect{x}+\Delta\vect{x}) + \vect{n'}_{\,\mathcal{I}} = \vect{y(\vect{x})} + \Delta\vect{y} \nonumber
\end{eqnarray}
and take the average Euclidean distance between them, resulting in:
\begin{eqnarray}
      D_{\textrm{th}}^2 &=& \mathbb{E}\left[ \, \Delta\vect{y}^\top\cdot\Delta\vect{y}  \,  \right] \nonumber \\
      &=& \mathbb{E}\left[  \,\,\,   \left|\Delta S + \vect{n'}_{\,\mathcal{I}} - \vect{n}_{\mathcal{I}} \right|^2  \,\,   \right]
      \label{distance_non_param_AppendixC}
\end{eqnarray}
where $\Delta S = S(\vect{x}+\Delta\vect{x}) - S(\vect{x})$ is the difference between the deterministic responses, and $\vect{n}_{\mathcal{I}}$ and $\vect{n'}_{\mathcal{I}}$ are different realizations of the \emph{inner} noise at the points $S(\vect{x})$ and $S(\vect{x}+\Delta\vect{x})$, respectively. Note that, as already said, Eq.~\ref{distance_non_param_AppendixC} means that when judging the difference between two stimuli the brain compares two noisy responses, $\vect{y}(\vect{x})$ and $\vect{y}(\vect{x})+\Delta{\vect{y}}$, that is, $S(\vect{x})+\vect{n}_{\,\mathcal{I}}$ and $S(\vect{x}+\Delta\vect{x}) + \vect{n'}_{\,\mathcal{I}}$.

As stated above, our proposal for noise estimation explained in section 3.1 consisted of finding the noise parameters that maximize the correlation between the experimental $D_{\textrm{exp}}$ and theoretical $D_{\textrm{th}}$ distances. As shown in Eq. (S7.3) and (S7.5) in \citep{Martinez18}, given $n$ experimental conditions to measure $n$ thresholds, the maximization of this correlation requires its derivative with regard to the parameters
of the model, and reduces to computing $\frac{\delta D_{\textrm{th}}^i}{\delta \theta}$, where $\theta$ are the noise parameters, and $D_{\textrm{th}}^i$ is the distance for the $i$-th experimental condition, with $i=1,\ldots,n$. According to Eq. (S7.6) in \citep{Martinez18}, these derivatives can be written as:
\begin{equation}
    \frac{\delta D_{\textrm{th}}^i}{\delta \theta} = \frac{1}{D_{\textrm{th}}^i} \cdot \left(\vect{y}(x^i+\Delta x^i) - \vect{y}(x^i) \right)^\top \cdot \left[ \frac{\delta \vect{y}(x^i+\Delta x^i)}{\delta \theta} - \frac{\delta \vect{y}(x^i)}{\delta \theta}\right]
\end{equation}
To compute the derivatives $\frac{\delta \vect{y}(\cdot)}{\delta \theta}$ we can apply the Taylor approximation to Eq. \ref{noisy_resp2} to get:
\begin{equation}
    \vect{y}(\vect{x}) = S(\vect{x}+\vect{n}_\varepsilon + \vect{n}_e) + \vect{n}_l \approx S(\vect{x}) + \nabla S \cdot \vect{n}_{\varepsilon} + \nabla S \cdot \vect{n}_e + \vect{n}_l
    \label{y_AppendixC}
\end{equation}

And now, for the case in which \emph{early} and \emph{late} noises are signal-dependent Gaussian-Poisson variables, plugging Eq. \ref{Gauss_Poisson_AppendixA} into Eq. \ref{y_AppendixC} we obtain:
\begin{equation}
    \vect{y}(\vect{x}) = S(\vect{x}) + \nabla S \cdot \vect{n}_{\varepsilon} + \nabla S \cdot \left( \alpha_e I + \beta_e\,\mathbb{D}_{|\vect{x}|^{1/2}} \right) \cdot \vect{n}_{G1} + \left( \alpha_l I + \beta_l\,\mathbb{D}_{|S(\vect{x}+\vect{n}_\varepsilon+\vect{n}_e)|^{1/2}} \right) \cdot \vect{n}_{G2}
\end{equation}
Applying again the Taylor approximation to the term $|S(\vect{x}+\vect{n}_\varepsilon+\vect{n}_e)|^{1/2}$ we finally get:
\begin{equation}
    \vect{y}(\vect{x}) = S + \nabla S \cdot \vect{n}^{\varepsilon} +
    \nabla S \cdot \left( \alpha_e I + \beta_e\,\mathbb{D}_{|\vect{x}|^{1/2}} \right) \cdot \vect{n}_{G1} +
    \left( \alpha_l I + \beta_l\,\mathbb{D}_{|S|^{1/2}+\frac{1}{2}|S|^{-1/2}\cdot \textrm{sign}(S) \cdot \nabla S \cdot (\vect{n}_\varepsilon + \vect{n}_e)} \right) \cdot \vect{n}_{G2}
\end{equation}
where for readability the function $S(\vect{x})$ has been named $S$, and the function $\textrm{sign}(\cdot)$ applies the sign of $(\cdot)$.

With this result for $\vect{y}(\vect{x})$ it is possible to obtain the derivatives wrt the \emph{early} $(\alpha_e, \beta_e)$ and \emph{late} $(\alpha_l, \beta_l)$ noise parameters that are needed to maximize the correlation between the experimental $D_{\textrm{exp}}$ and theoretical $D_{\textrm{th}}$ distances:
\begin{eqnarray}
    \frac{\delta \vect{y}(\vect{x})}{\delta \alpha_e} &=& \nabla S \cdot I \cdot n_{G1} +
    \left( \beta_l\,\mathbb{D}_{\frac{1}{2}|S|^{-1/2}\cdot \textrm{sign}(S) \cdot \nabla S \cdot n_{G1}}\right) \cdot n_{G2}\\
    \frac{\delta \vect{y}(\vect{x})}{\delta \beta_e} &=& \nabla S \cdot \mathbb{D}_{|\vect{x}|^{1/2}} \cdot n_{G1} +
    \left( \beta_l\,\mathbb{D}_{\frac{1}{2}|S|^{-1/2}\cdot \textrm{sign}(S) \cdot \nabla S \cdot \mathbb{D}_{|\vect{x}|^{-1/2}} \cdot n_{G1}}\right) \cdot n_{G2}\nonumber\\
    \frac{\delta \vect{y}(\vect{x})}{\delta \alpha_l} &=& n_{G2} \nonumber \\
    \frac{\delta \vect{y}(\vect{x})}{\delta \beta_l} &=& \left( \mathbb{D}_{|S|^{1/2}+\frac{1}{2}|S|^{-1/2}\cdot \textrm{sign}(S) \cdot \nabla S \cdot \left( \alpha_e I + \beta_e\,\mathbb{D}_{|\vect{x}|^{1/2}} \right) \cdot \vect{n}_{G1}}\right) \cdot n_{G2} \nonumber
\end{eqnarray}

As can be seen from these last equations, in this non-parametric case the derivatives wrt the \emph{early} and \emph{late} noise parameters do not involve matrix inversion, and hence they are easy to compute. This allows a practical optimization of the noise parameters to maximize the correlation between the experimental and theoretical distances, no matter the complexity of the noise, whenever $\nabla S$ is easy to compute.

\section{Appendix C: Equivalence of parametric and non-parametric distances}

In this appendix we show that the considered distances, Eq.~\ref{metric_noise1} and Eq.~\ref{distance_non_param}, are equally valid to get the corresponding estimation of the noise.
First we show that the distances have an interesting anisotropic behavior depending on the noise: perceptual distance is markedly different for distortions in different directions $\Delta\vect{x}$.
The noise-dependent anisotropy is obvious in Eq.~\ref{metric_noise1} because the metric depends on the covariance of the noise, but this is less obvious in the non-parametric Eq.~\ref{distance_non_param}.

The first part of this Appendix analytically shows that the non-parametric distance displays this anisotropy.
Afterwards, we numerically show that the distance computed in both ways can be linearly related.
As a result, since correlation is independent of a linear transformation applied to one of the axis, both definitions of $D_{\textrm{th}}$ are equivalent for our purposes.

\noindent \textbf{Analytical part: non-parametric distance is anisotropic}.
The general expression for the average Euclidean distance in the non-parametric case, Eq. \ref{distance_non_param_AppendixC}, $D_{\textrm{th}}^2 = \mathbb{E}\left[  \,\,\,   \left|\Delta S + \vect{n'}_{\,\mathcal{I}} - \vect{n}_{\mathcal{I}} \right|^2  \,\,   \right]$, can be written as:
\begin{equation}
    D_{\textrm{th}}^2 = \mathbb{E}\left[\left( \Delta S + \left(\vect{n'}_{\,\mathcal{I}} - \vect{n}_{\mathcal{I}}\right) \right)^\top \cdot \left( \Delta S + \left(\vect{n'}_{\,\mathcal{I}} - \vect{n}_{\mathcal{I}}\right) \right) \right]
\end{equation}
which can be expanded as follows:
\begin{eqnarray}
    D_{\textrm{th}}^2 &=& \mathbb{E}\left[\Delta S^\top \cdot \Delta S + \Delta S^\top \cdot \left(\vect{n'}_{\,\mathcal{I}} - \vect{n}_{\mathcal{I}}\right) +
    \left(\vect{n'}_{\,\mathcal{I}} - \vect{n}_{\mathcal{I}}\right)^\top \cdot \Delta S +
    \left(\vect{n'}_{\,\mathcal{I}} - \vect{n}_{\mathcal{I}}\right)^\top \cdot \left(\vect{n'}_{\,\mathcal{I}} - \vect{n}_{\mathcal{I}}\right)\right]\nonumber \\
    &=& \Delta S^\top \cdot \Delta S +
    2 \, \mathbb{E}\left[\Delta S^\top \cdot \left(\vect{n'}_{\,\mathcal{I}} - \vect{n}_{\mathcal{I}}\right)\right] +
    \mathbb{E}\left[\vect{n'}_{\,\mathcal{I}}^\top \cdot \vect{n'}_{\mathcal{I}}\right] -
    2\,\mathbb{E}\left[\vect{n'}_{\,\mathcal{I}}^\top \cdot \vect{n}_{\mathcal{I}}\right] +
    \mathbb{E}\left[\vect{n}_{\,\mathcal{I}}^\top \cdot \vect{n}_{\mathcal{I}}\right]\nonumber \\
    &=& \left|\Delta S\right|^2 +
    \left|\vect{n}_{\,\mathcal{I}}\right|^2 +
    \left|\vect{n'}_{\,\mathcal{I}}\right|^2 -
    2\,\mathbb{E}\left[\vect{n'}_{\,\mathcal{I}}^\top \cdot \vect{n}_{\mathcal{I}}\right] +
    2 \, \mathbb{E}\left[\Delta S^\top \cdot \left(\vect{n'}_{\,\mathcal{I}} - \vect{n}_{\mathcal{I}}\right)\right]
\end{eqnarray}
where the last term vanishes because it compares a deterministic component and a random component. When $\Delta S = 0$, $\vect{n'}_{\,\mathcal{I}} = \vect{n}_{\,\mathcal{I}}$, and then, $2\,\mathbb{E}\left[\vect{n'}_{\,\mathcal{I}}^\top \cdot \vect{n}_{\mathcal{I}}\right] = 2 \left|\vect{n}_{\,\mathcal{I}}\right|^2$, so $D_{\textrm{th}}^2 = 0$. When $\Delta S \neq 0$, as $\vect{n}_{\,\mathcal{I}}$ and $\vect{n'}_{\,\mathcal{I}}$ are decorrelated, $\mathbb{E}\left[\vect{n'}_{\,\mathcal{I}}^\top \cdot \vect{n}_{\mathcal{I}}\right] = 0$, so we get:
\begin{equation}
     D_{\textrm{th}}^2 = \left|\Delta S\right|^2 +
    \left|\vect{n}_{\,\mathcal{I}}\right|^2 +
    \left|\vect{n'}_{\,\mathcal{I}}\right|^2
\end{equation}
Note that, in this expression, $|\Delta S|$ and $|\vect{n}_{\,\mathcal{I}}|$ are constants for a given $\vect{x}$. However, in general, the energy $\left|\vect{n'}_{\,\mathcal{I}}\right|^2$ depends on $\Delta \vect{x}$ (or $\Delta S$). This dependence is what generates the anisotropic behavior of the distance. For instance, if the \emph{inner} noise has a Poisson component, the direction of $\Delta S$ matters: increasing the response in a certain direction may increase the energy, while going in the opposite direction (reducing the response) may reduce the energy of the noise.

\begin{figure}[t!]
\begin{center}
\includegraphics[width=0.7\textwidth]{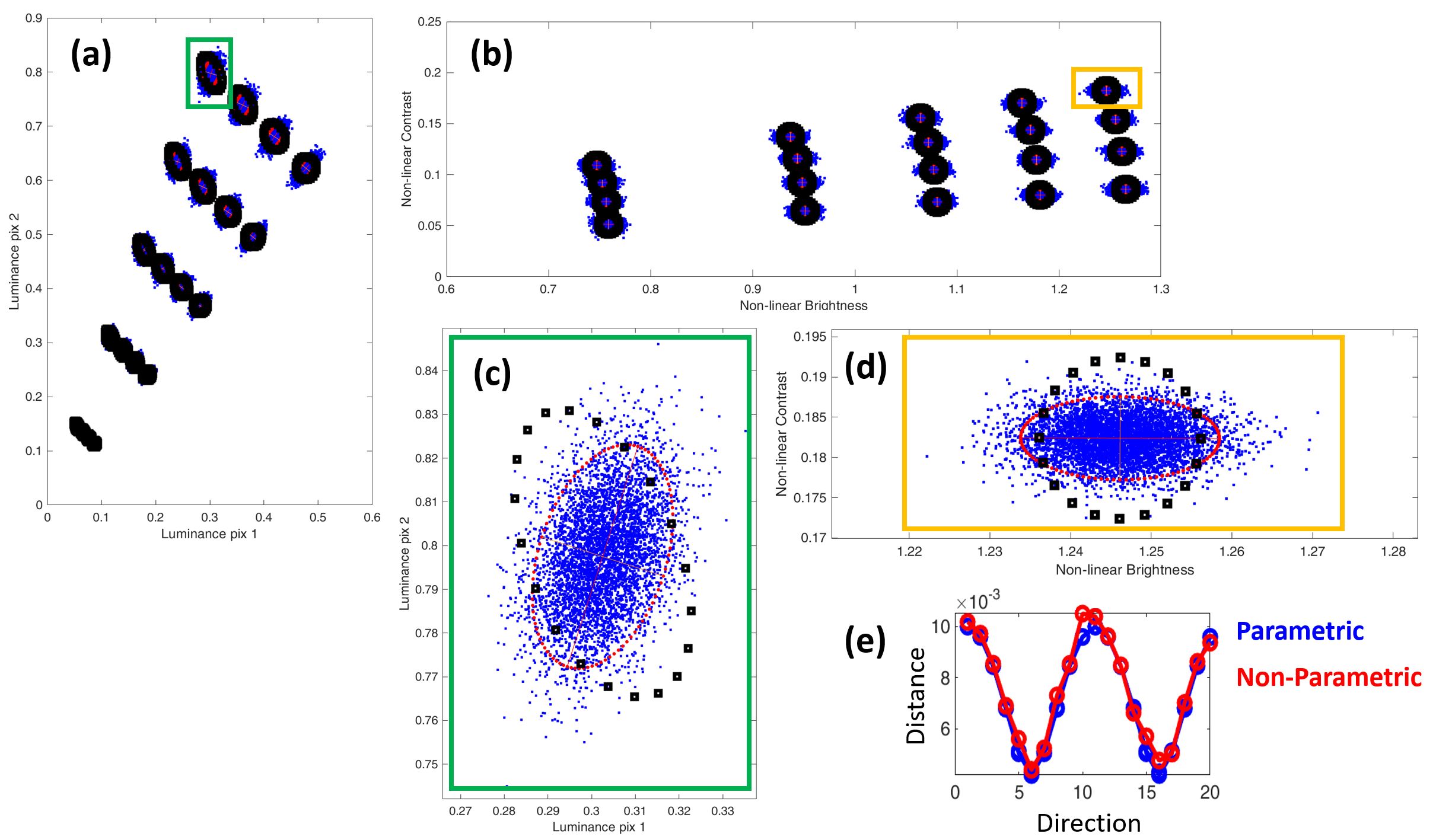}\\
\includegraphics[width=0.7\textwidth]{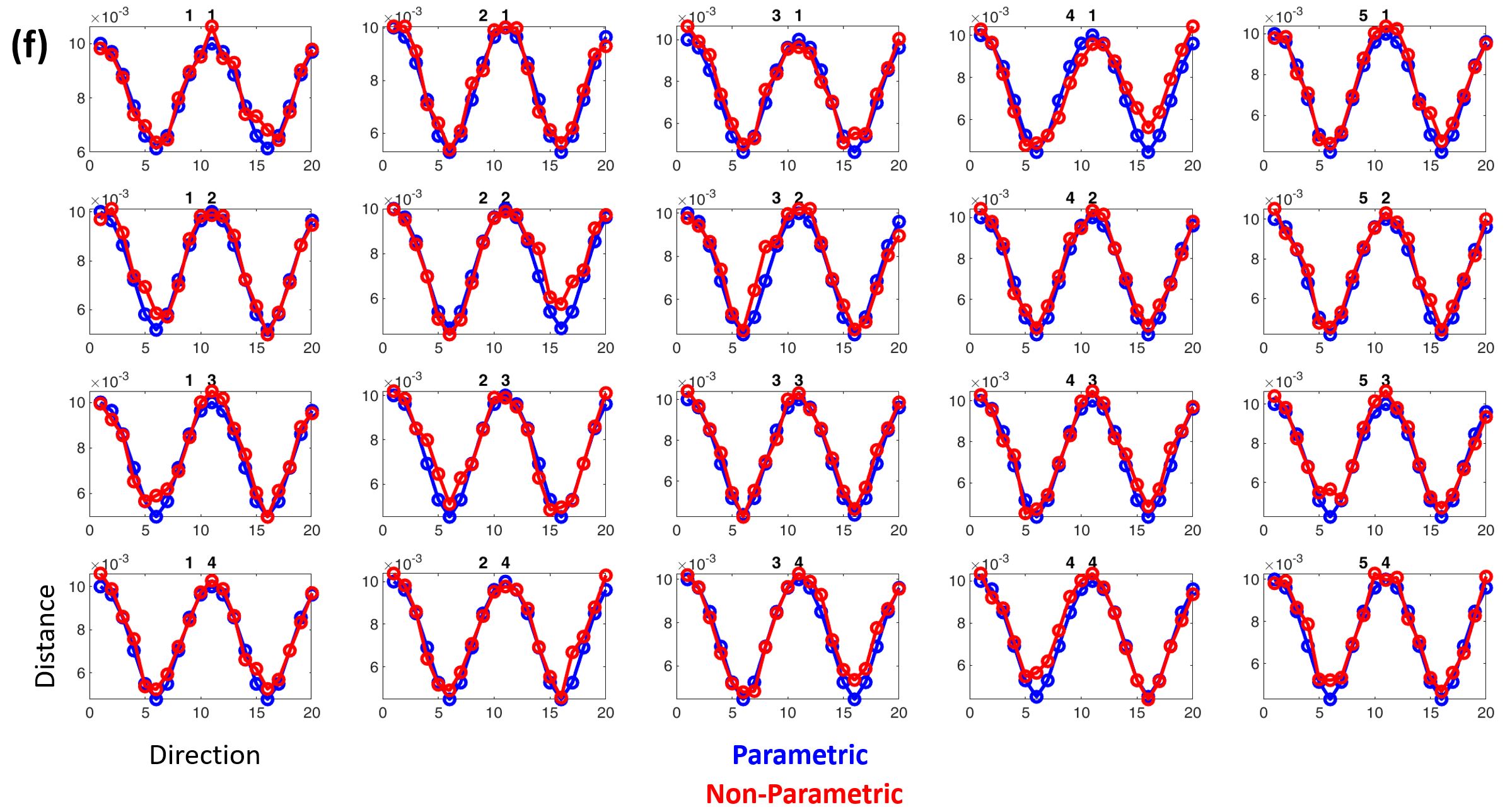}\\
       [-0.0cm]
       \caption{\small{\textbf{Equivalence between parametric and non-parametric distances.} 
       Plots~(a) and~(b) show the $4 \times 5$ locations (4 contrasts times 5 luminances)  considered in the space of 2-pixel images and in the inner representation of the illustrative model introduced in Section 2.3.
       The noisy images and responses are represented by the random samples in blue. Plots~(c) and~(d) zoom the highlighted rectangles in green and orange. In each location in the inner representation we computed distances from the central point in the directions indicated by the black squares. 
       The plot~(e) represents the distances from the central point in plot (d) in the different directions computed through the parametric and non-parametric methods, Eqs.~\ref{metric_noise1} and~\ref{distance_non_param}, in blue and red respectively. The plots in panel~(f) show the same result at the other considered points all over the image space.
      }}
      \label{equivalence}
      \end{center}
\end{figure}

\noindent \textbf{Numerical illustration: non-parametric and parametric distances are linearly related.} 
In this section we use the simplified vision model introduced in Section 2.3 (on 2-pixel images) to illustrate explicitly how the non-parametric distance of Eq.~\ref{distance_non_param} follows the trends of the parametric metric in Eq.~\ref{metric_noise1}.
Using the model implemented here\footnote{\texttt{http://isp.uv.es/code/visioncolor/noise.html}} we considered $4 \times 5$ locations (4 contrasts times 5 luminances) in the space of 2-pixel images, and we generated 5000 noisy responses for each of these images. See the noisy samples and responses in Fig.~\ref{equivalence}. Then, we considered 20 points at constant Euclidean distance from each average response in equidistant directions along the Euclidean sphere, e.g. the black squares in the zoom Fig.~\ref{equivalence}.d. 
These points are convenient to point out the anisotropy of the measures and their eventual equivalence.

The metric defined by the noisy samples implies that the locus of equidistant points from the center is the ellipsoid highlighted in red. 
As a result, the perceptual metric is anisotropic: starting from the vertical direction in Fig.~\ref{equivalence}.d, the distance is maximum and it goes down and up in a periodic way, as given by the blue line in plot~\ref{equivalence}.e, computed according to the parametric distance. The distance of each black square was also computed from the expected value in Eq.~\ref{distance_non_param} leading to the red curve in 
plot~\ref{equivalence}.e.
The plots in the panel~(f) show that the equivalence between the two measures holds all over the image space with the same linear relation between the parametric and non-parameteric measures. All the curves in red were obtained from the non-parametric distances and the same linear relation with the parametric distance.

In summary, the (simpler to compute) non-parametric distance based on the average over the noisy samples is equivalent, i.e. leads to the same correlation, as the parametric distance which requires the inversion of huge matrices in the optimization.
 

\section{Appendix D: Noise from thresholds for individual observers}

\begin{figure}[h!]
\begin{center}
\includegraphics[width=0.85\textwidth]{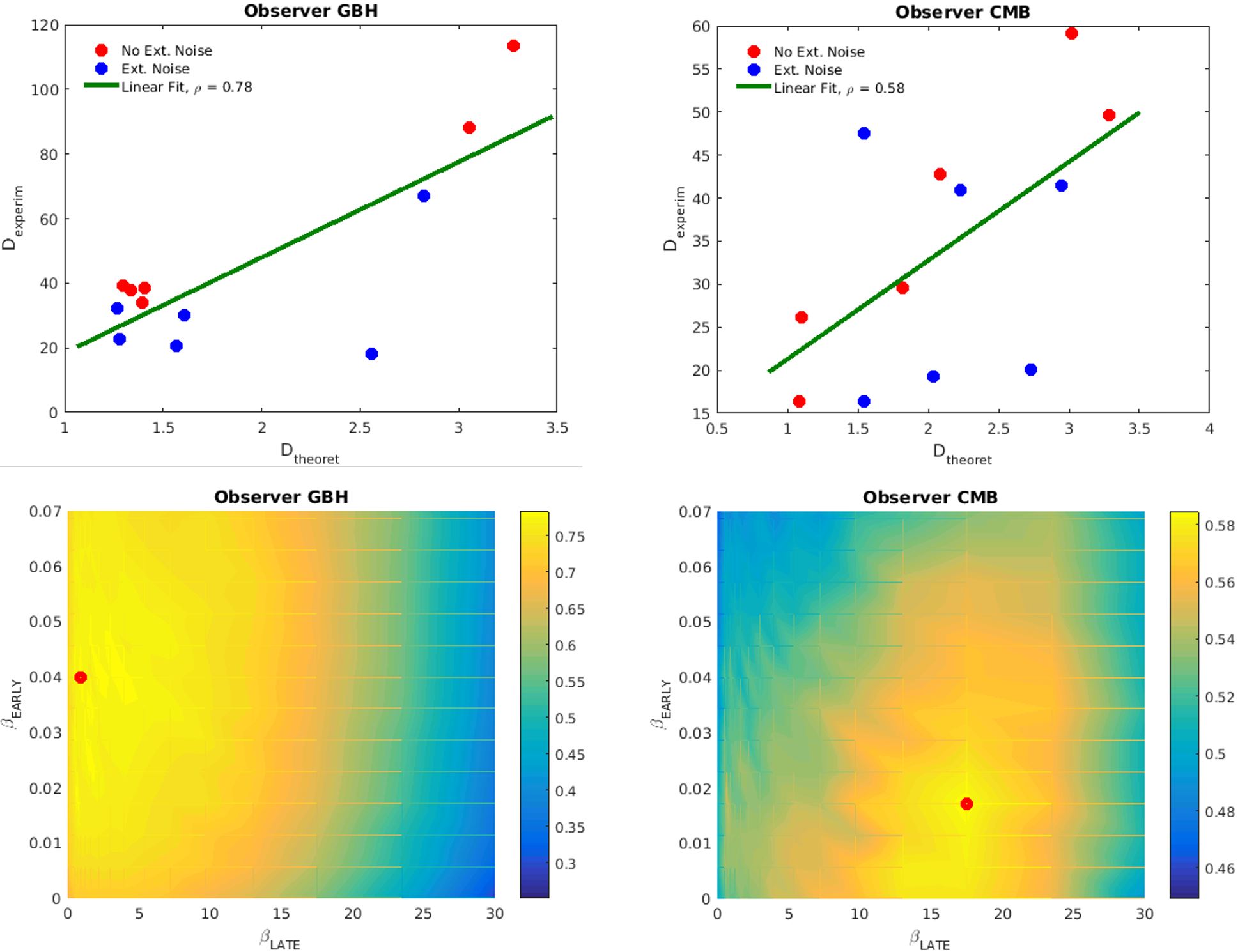}\\[-0.0cm]
\caption{\small{\textbf{Early and late noise parameters from 
        threshold data for each individual observer.} 
        The top plots represent the best (maximum correlation) linear fits between the theoretical distance (that depends on the noise sources) and the experimental distance (based on the thresholds) for the considered observers.
        The surfaces at the bottom display the correlation 
        between the theoretical distance and the experimental distance depending on the noise parameters.
        The optimal noise parameters, here highlighted in red, are those that maximize the correlation for each observer 
        ($\beta_e = 0.04$, $\beta_l = 0.91$) for GBH and ($\beta_e = 0.02$, $\beta_l = 17.5$) for CMB.}}
        \label{main_result1_separated}
        \end{center}
  \end{figure}

The surfaces in Figure~\ref{main_result1_separated}
show the correlation between the threshold-based experimental distances and the 
theoretical distances based on pure Poisson early and late noise for each individual observer.
In this case the data of each observer is considered separately, so we obtain different optima (highlighted in red) for each observer. The scatter plots on top show the predictions with better linear correlation.  

In this threshold-only method separation of data per observer does not improve the optimization problem: similarly to Fig.~\ref{main_result1} the correlation surfaces display large plateaus of almost constant values thus indicating uncertainty in the optima.
Moreover, observers display quite different results. In particular observer GBH seems to drive the global result in Fig.~\ref{main_result1} because the data from observer CMB displays small correlation anyway and has small impact in the global correlation.

This uncertainty and disparity is in contrast with the narrow minima and consistency between observers found in the method that considers all the data in the psychometric functions (Fig.~\ref{main_result2}).
This suggests that the consideration of the full psychometric functions properly constrain the problem towards  more accurate and consistent solutions.
        
\section{Appendix E: Role of early and late noise in detection and discrimination}

To understand better the implications of the estimated noise parameters, we simulated the model with those parameters and analysed the representations for a reduced case similar to our two-pixel toy example from section 2.
For this we generate noisy responses for gratings of 2 c/deg of different contrast (0.01, 0.15 and 0.7) and different average luminance (15, 50, 100 $cd/m^2$) using the model in Eq.~\ref{modular} and the results of the Poisson noise sources obtained with the full-psychometric method. 

\begin{figure}[h!]
  \begin{center}
  STIMULUS SPACE\\
  \includegraphics[width=0.63\textwidth]{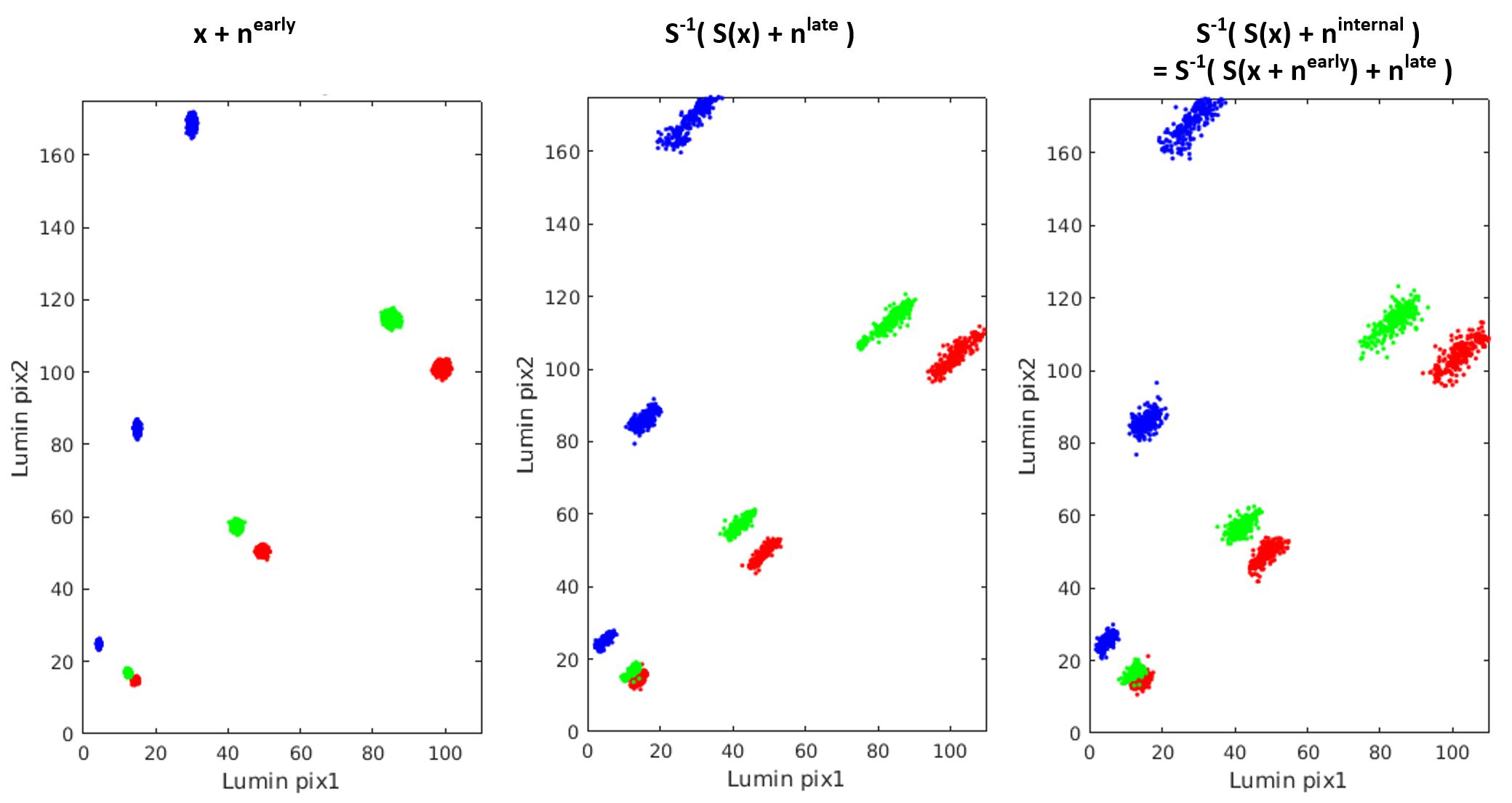}\\  
  INNER (LATE) REPRESENTATION\\
\includegraphics[width=0.63\textwidth]{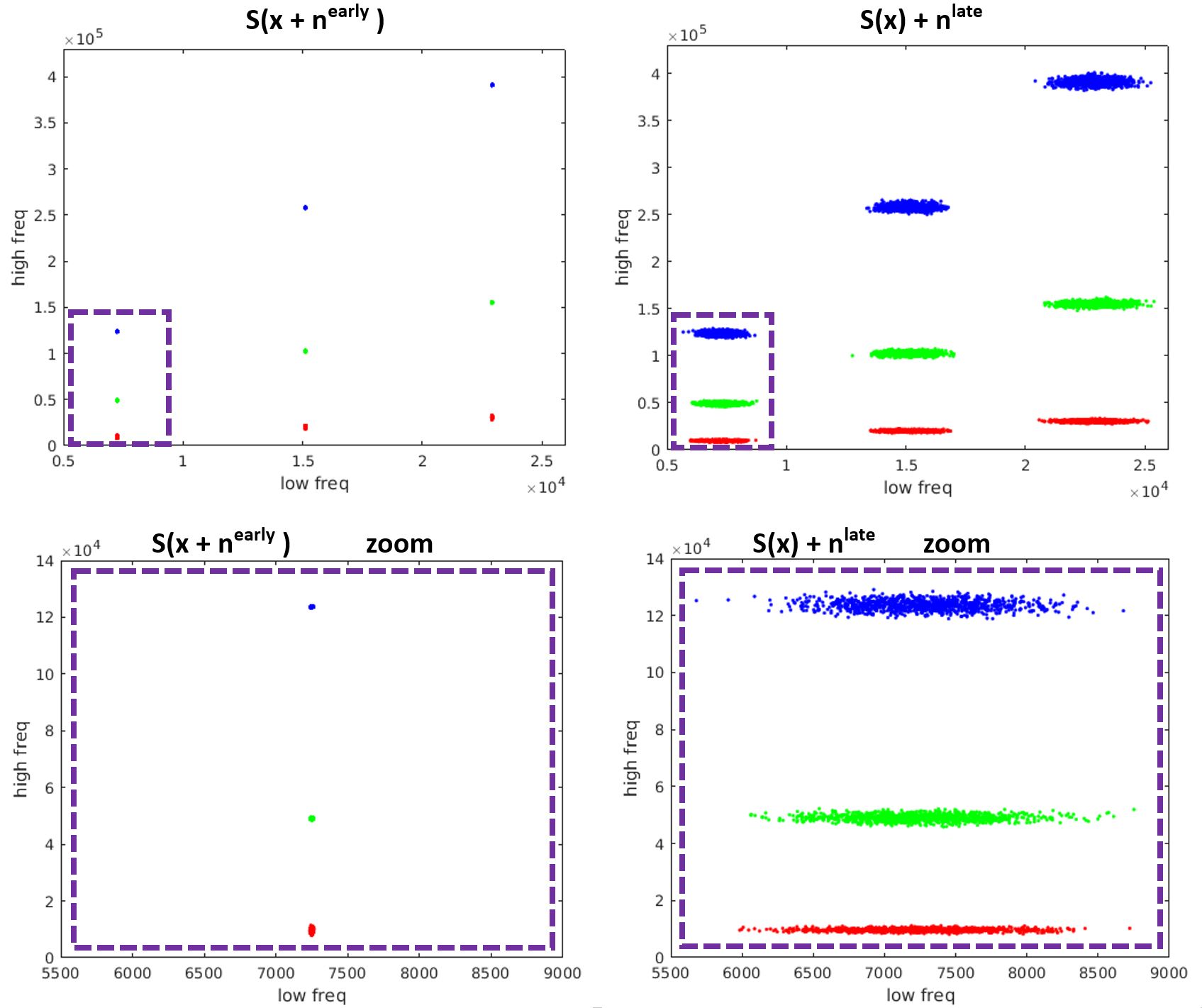}\\    
        [-0.0cm]
        \caption{\small{\textbf{Gratings corrupted by early and late noise with noise parameters as estimated in section 4.2}. We show samples in 2-d projections of the stimulus space (top row) and the late representation (middle and bottom rows). 
        The signal in the input representation corresponds to the brightest and the darkest locations of a 2cpd grating for different values of luminance and contrasts. 
        Here, red, green, and blue clusters correspond to progressively higher values of contrast [0.01, 0.15, 0.7]. And clusters progressively further away from the origin in the diagonal direction of the stumulus space correspond to images with higher average luminance, with values [15, 50, 100] $cd/m^2$.
        The middle row shows the samples in the 2-dimensional representation along the zero-frequency dimension (horizontal axis) and along a frequency of 2 c/deg (vertical axis). 
        The rectangles in dashed style highlight the region of low-luminance gratings of different contrast. The bottom row zooms-in this region.
        }}
        \label{simul_noise1}
        \end{center}
  \end{figure}

Figure~\ref{simul_noise1} shows the noisy signals with different contributions of the noise, at the  two levels of representation. Computations are done with a full scale model that works with $256\times256$ images. 
However, these two-dimensional projections are obtained by selecting two sensors of the image representation and two sensors of the late representation.
For illustrative purposes, we took for the input the ones corresponding to the darkest and lightest locations of the gratings, and from the late representation a wavelet unit from the low-frequency residual and another tuned to the 2 cycles/deg band. 
In this way, the meaning of the representations in Fig.~\ref{simul_noise1} is qualitatively similar to the representations in toy scenario presented in Figs.~\ref{twopixel_basic} and~\ref{twopixel_external}: 
the input spaces are completely equivalent, and in the projection that we consider here the horizontal axis also represents brightness and the vertical axis also represents contrast. 

In this context, the clusters of the samples corrupted by the early noise in the input domain (top-left plot) display the properties of the assumed early Poisson noise: note that (1) the width (or variance) of the clusters increases with the luminance, and (2) the ellipsoidal clusters are aligned with the axes because we assumed no correlation.
Similarly, the late noise in the late representation (middle-right and bottom-right plots) also displays the Poisson properties: the variance increases with the average and there is no correlation between their uncertainty.
Interestingly, when the late noise is transformed back into the early representation (e.g. top-center plot) there is a strong correlation in the uncertainty of the different photoreceptors.
The same is true in the case of the early noise in the late representation (its covariance is non-diagonal), although it is hard to see given the small scale of the early noise (middle-left and bottom-left plots).

The interesting interaction between the Poisson nature of the noise sources and the nonlinearities of the model implies that in the late representation (where decisions are made), the different sources of noise (early or late) determine the performance in different conditions (detection or discrimination).
Note that the saturating nonlinearity in the contrast response implies that the clusters of early noise get compressed for progressively bigger contrasts (see that the red cluster is wider than the blue in the bottom-left plot).
On the contrary, the Poisson nature of the late noise implies that the uncertainty due to this noise for the high contrast signals is bigger than the variability introduced for the responses to low contrast gratings (blue cluster is wider than the red in the bottom-right plot).  

  \begin{figure}[t!]
  \begin{center}
\includegraphics[width=0.9\textwidth]{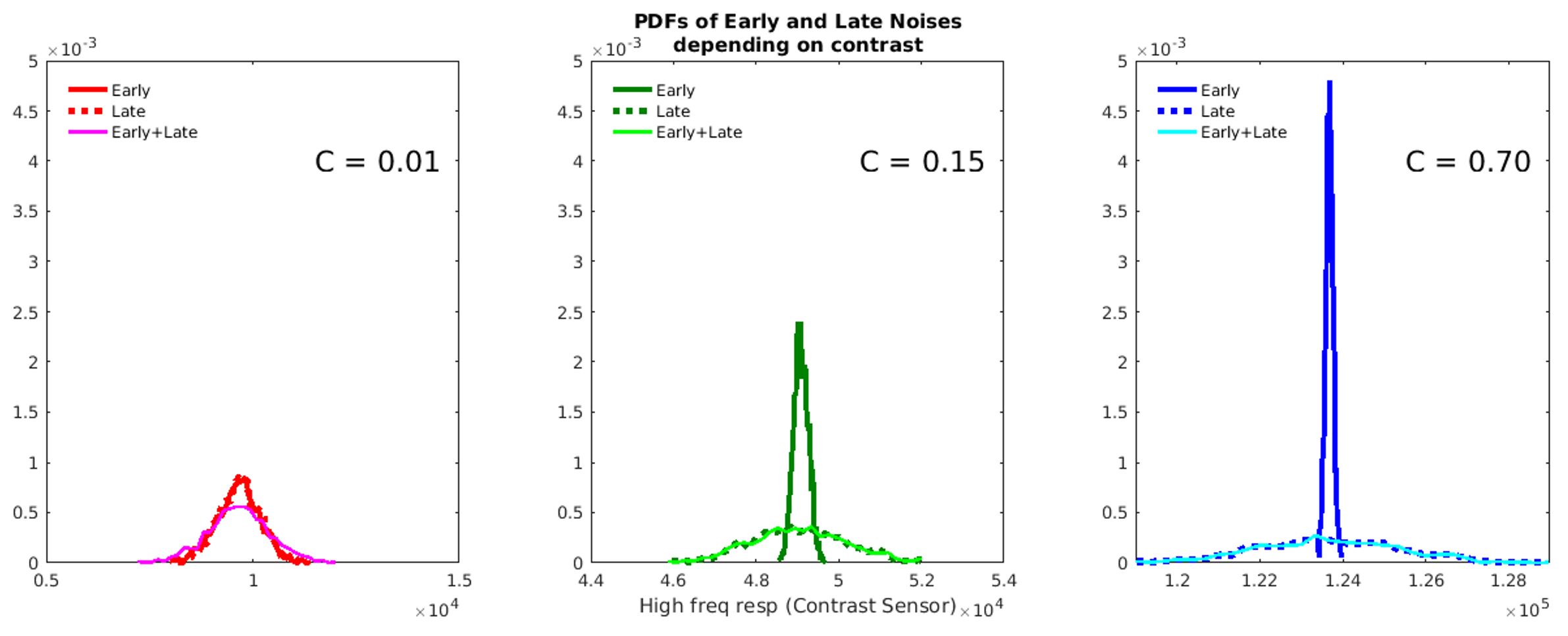}\\    
        [-0.0cm]
    \caption{\small{\textbf{Marginal PDFs of the early, late, and (early+late) noise sources in the inner representation}.
        These PDFs are computed from the samples of the 2 cycle/deg sinusoidal gratings of different contrast (with average luminance of 15 $cd/m^2$) shown in the bottom row of Fig.~\ref{simul_noise1}. The marginals are taken in the direction of the sensor tuned to 2 cycl/deg (vertical axis in Fig.~\ref{simul_noise1}). 
        }}
        \label{simul_noise2}
        \end{center}
  \end{figure}

This is even more clearly represented in Figure~\ref{simul_noise2}, which shows the marginal PDFs for three cases in the bottom plots of Fig. \ref{simul_noise1}. Here the PDFs of the early and late noise are basically the same for low contrast (C = 0.01), but when the contrast is increased, the PDFs of the early noise are squeezed (solid curves) while the PDFs of the late noise widen (dashed curves). 
For low contrast gratings the total inner noise is basically determined only by early noise (and this effect would be even bigger for close-to-zero contrast used in actual detection experiments), while for high contrast gratings the inner noise is determined by late noise.

This explains why the consideration of one noise source or the other is important to reproduce all experimental psychometric functions in Figure~\ref{main_result2}, and the different role of early and late noise in detection and discrimination.

\section{Appendix F: Psychophysical versus physiological estimates of early noise}


The results and discussion in this Appendix clarifies why it is sensible to have one order of magnitude less noise in a psychophysical model rather than in a physiological model of the retina. 

\begin{figure}[t!]
\begin{center}
   \vspace{-0.5cm}
   \includegraphics[width=0.4\textwidth]{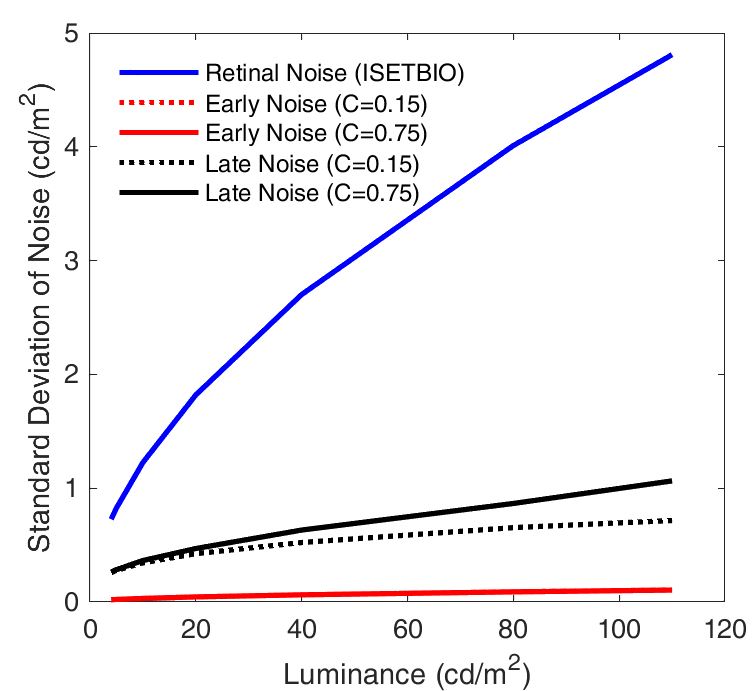}\\[0.25cm]
   \includegraphics[width=1\textwidth]{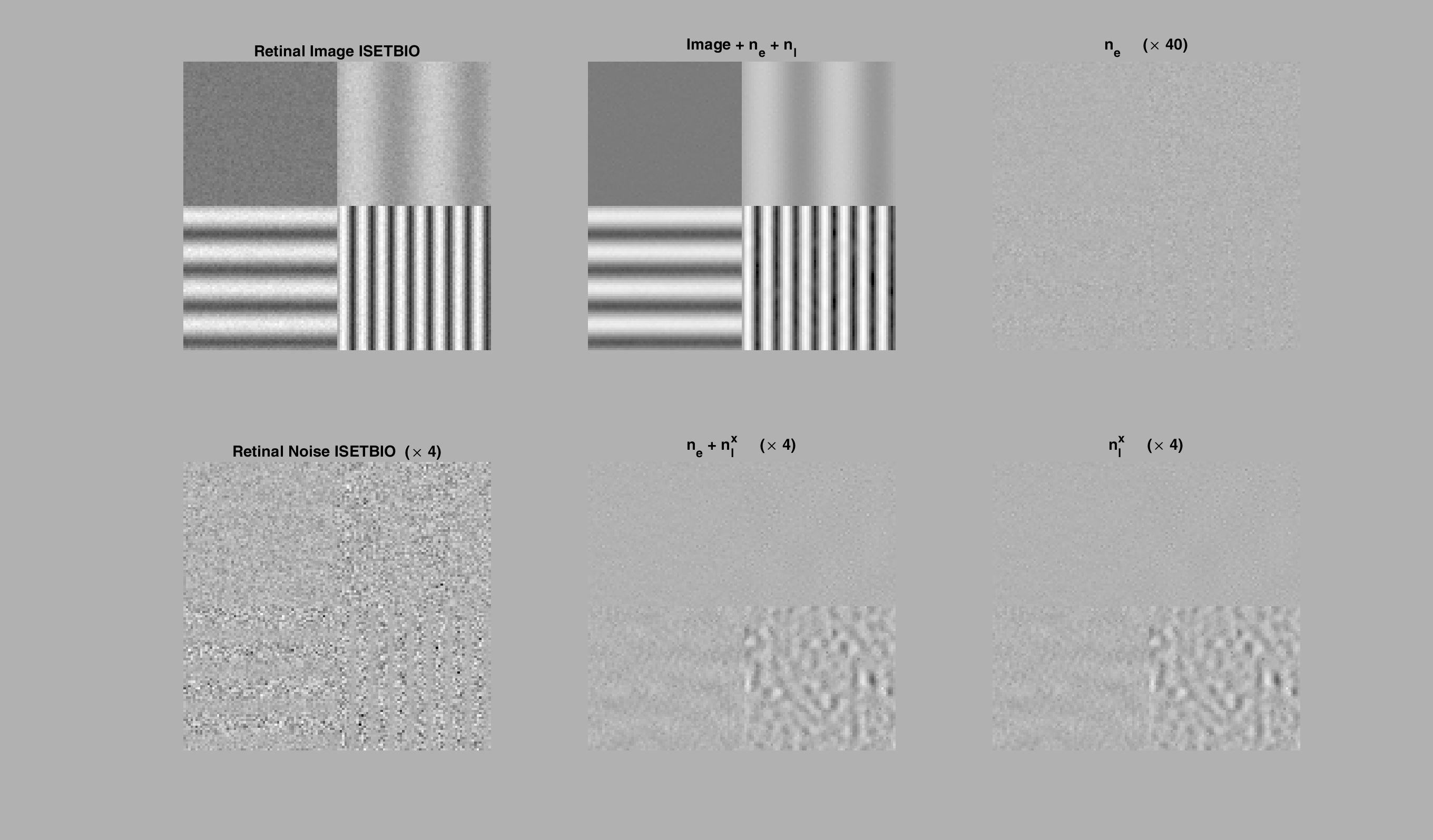}\\
   \caption{\small{\textbf{Comparison of the physiological (retinal) noise estimated in Appendix E following~\blue{(Brainard \& Wandell, 2020)}, and the psychophysical (Early and Late) estimations proposed here}.
   \emph{Top panel:} shows the standard deviation of the early and the late noises as a function of the luminance and contrast of a 4 cpd sinusoid.
   As expected from the Weber law, noise increases with luminance. 
   Moreover, according with masking, the late noise increases with contrast, which is not the case for early noise.
   The deviation of the retinal noise (see Appendix E) is plotted in the same units as a reference.
   \emph{Bottom panel:} shows an illustrative stimulus calibrated in luminance with the noise of both models inverted back into the image space.
   Luminance of the flat square is 30~$cd/m^2$. Average luminance of the sinusoids is 60~$cd/m^2$. Frequencies are 2, 4 and 8 cpd, and contrasts are 0.25, 0.75, and 0.9 respectively.}}
   \label{noisesPoisson1}
\end{center}
\end{figure}

Fig.~\ref{noisesPoisson1} (top panel) shows the standard deviation of our estimated early and late noise sources on top of a sinusoid for different average luminance and different contrasts. The equivalent ISETBio noise in $cd/m^2$ is plotted on top for useful reference.
Fig.~\ref{noisesPoisson1} (bottom panel) shows sample images with ISETBIO noise and with the inner noise estimated by us, both back in the spatial domain.

If noise controls the discriminability, it should be just noticeable (or almost invisible) for the average observer.
Clearly, that is not the case for the physiological noise in the retina: ISETBio noise is too big and clearly visible. This indicates the presence of down-stream mechanisms
to remove this uncertainty (motion compensation + evidence accumulation over time + spatial low-pass filtering...).
Interestingly our noise estimate is almost invisible on top of the background pattern. In order to explore its nature and assess its early/late components we show scaled versions of the noises on top of a flat background of 60~$cd/m^2$.
While the Poisson early noise (trivially) inherits the spatial structure of the luminance, the late noise reveals a contrast/frequency dependence due to the inner working of the nonlinear wavelet filters of the deterministic model. Note that more noise is allocated in the high contrast regions. This is consistent with the results in Fig.~\ref{simul_noise1}-(bottom right) and Fig.~\ref{simul_noise2}-right, that highlight the increase of the variance of late noise with contrast. 
Moreover, note also in Fig.~\ref{noisesPoisson1} that the frequency of the noise depends on the frequency of the background pattern because more active sensors are more affected by the late noise.
A final factor that affects the magnitude of the late noise in the input domain is the inverse of the CSF that enhances the amplitude of the (less visible) high-frequency noise.

\vspace{0.25cm}

Below we described how we used ISETBIO~\citep{Cotaris1,Cotaris2} to get a reasonable physiological estimation of retinal noise.
ISETBio allows the definition of a stimulus in physical units, e.g. an achromatic stimulus (flat spectral radiance) of certain lumianance in $cd/m^2$, and subtending certain degrees in the visual field.
The software incorporates accurate models of the human optics and of retina photoreceptors, so it computes both the blurred image at the retina and models the isomerization process so that the retinal image is transformed in noisy photocurrents at the L, M and S sensors.
This happens dynamically and incorporates fixation and microsaccade motion.

One can use this accurate model to estimate the noise added to the input in $cd/m^2$ units by doing two things: (1)~modeling the relation between input luminance and brightness expressed in terms of cone photocurrents, and (2)~computing the noise in the brightness domain and transforming it back to luminance using the inverse of the luminance-brightness relation.
In particular, we used the demo 
\verb!t_dynamicStimulusToPhotocurrent.m! with default parameters. We just modified the function to have access to intermediate results or switch on/off the eye motion model when necessary, but we did not change the default choices.

\begin{figure}[t!]
  \begin{center}
        \includegraphics[width=0.95\textwidth]{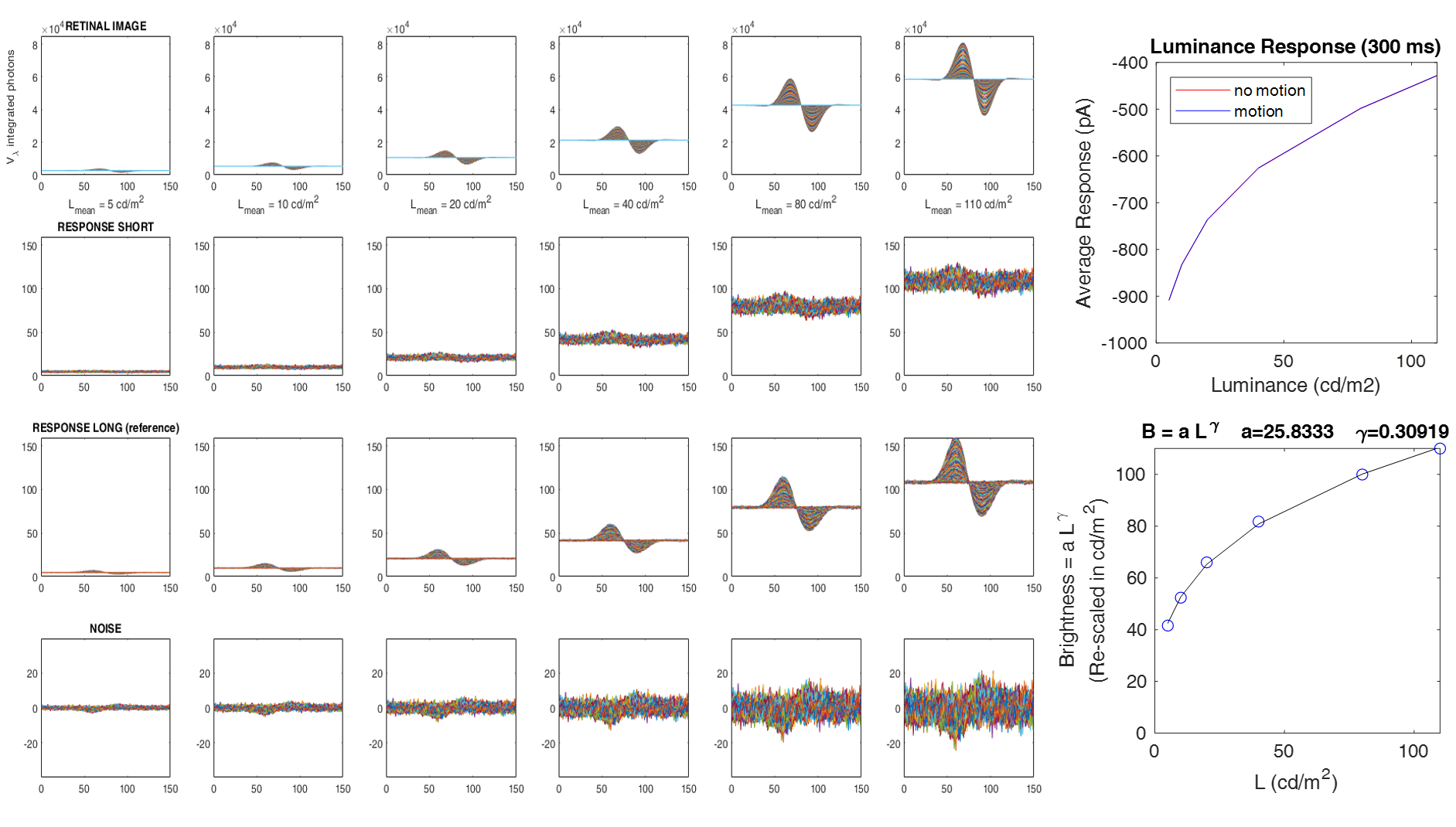}\\[-0.5cm]
        \caption{\small{\textbf{Procedure to estimate retinal noise in $cd/m^2$ using ISETBio.} Top row represents a series of achromatic Gabors of 2cpd of fixed contrast with controlled luminance. Top-right plot displays the nonlinear relation between input luminance and L+M photocurrents in $pA$ (with 200 msec  exposure time). Bottom-right plot displays the same relation where the vertical axis has been re-scaled to have values in the luminance range. The plot also displays the parameters of a exponential luminance-brightness fit. 
        The responses of the ISETBio retina corresponding to the stimuli in the 1st row, using either a short exposure time (hence noisy) and averaged over a long exposure time (hence de-noised) are shown in the 2nd and 3rd rows respectively. 
        The luminance-brightness fit has been used to express the responses of the 2nd and 3rd rows (given in $pA$ by ISETBio) in luminance units ($cd/m^2$). The subtraction of the 2nd and 3rd rows leads to the noises estimated in the 4th row.}}
        \label{fig_isetbio}
        \end{center}
  \end{figure}

The estimation of the luminance-brightness relation is easy to do assuming that the brightness perception is driven by the sum of the responses of the L and M cones, as is usual in color vision models~\citep{Fairchild13}. By doing so, one can build a series of stimli of different luminance, see the top row of Fig.~\ref{fig_isetbio}, and register the L+M responses in $pA$. 
In order to compare the spatial description of the stimuli and the photoreceptor mosaic We spatially interpolated the L and M mosaics using linear interpolation (i.e. we introduced minimal spatial blur). In particular, we didnt assume the wider spatial summation happening down stream in retinal ganglion cells or LGN cells. 
We used Gabors of 2 cpd with different average luminance and fixed contrast and put them through the model to empirically derive a luminance-photocurrent function. 
The responses integrated over short periods of time (e.g. 200 msec) display a substantial amount of noise but the average value of L+M in $pA$ can be used to derive a simple input-output curve that can be interpreted as a luminance-brightness response.
The nonlinear plots at the right of Fig.~\ref{fig_isetbio} display such curves (the top one in $pA$ and the bottom one re-scaled to have values in the range of the input luminance). We fitted a conventional exponential function to transform the input luminance into this \emph{brightness}: see the parameters of the fit in the bottom-right figure.
This function can be inverted in order to transform photocurrents into luminance values. 
This inverse is the one that allows us to represent the noisy responses obtained using short exposure times (200 msec) in a luminance scale (as for instance the second row of responses to the stimuli shown in the first row). 

We explicitly checked that the ocular motion has no effect in this estimation. Consistently with the reports in~\citep{Kelly79}, the speed induced by the ocular motion in ISETBio is about 0.1 deg/sec, which, in 200 msec just implies a spatial displacement of 0.02 degrees, which is small to induce a major luminance change in the 2cpd Gabors used in our experiment. Therefore the luminance-photocurrent curves do not change if we switch off the ocular motion in the model.

We estimated the noise in the beightness domain by subtracting this noisy L+M signal integrated over a short period of time (200 msec) from the equivalent signal averaged over a large exposure time to "artificially" remove all the noise in the response. 
In this second long-exposure-time case (where we integrated over 8 seconds) we switch off the ocular motion in the model so that the noise removal does not come from changes in the spatial structure of the input.
The temporally-denoised responses in luminance units can be seen in the third row of Fig.~\ref{fig_isetbio}. The subtraction of the clean response (3rd row) from the noisy responses (2nd row) leads to the noise realizations shown in the 4th row of the figure.
In this row that displays the physiological noise for different input luminance it is obvious that the variance of this noise increases with the input luminance.

Measuring the standard deviation of the noise in this series of realizations in the 4th row one can obtain the result shown in Fig.~\ref{noisesPoisson1}: according to ISETBio the standard deviation of the noise at the retina depends on the luminance $L$ as: $\sigma_{\textrm{retina}} \approx 0.33*L^{0.57} \,\, (cd/m^2)$.

\section{Appendix G: Readout metric and noise at the decision stage}

When there is noise at the decision stage or when the read out is not Euclidean, the proposed  Eq.~\ref{metric_noise1} changes as outlined below, but the relevance of the noise at the input to determine the scale of the different components of the noise mentioned above is still the same.

On the one hand, univariate noise at the decision stage could be introduced at the distance variable: $D_{\textrm{th}}^{\textrm{noisy}} = D_{\textrm{th}} + n_{\textrm{decision}}$. But this is equivalent to an extra expansion of the uncertainty in the direction of stimulus modification, whatever this direction is, i.e. an extra isotropic noise. This isotropic noise is equivalent to including an extra diagonal term in the covariance of the inner noise: $\Sigma_{\mathcal{I}} + k^2 \sigma^2 I$. Again, the scale factor $k$ of this isotropic noise could not be obtained with no external nor early noise.

On the other hand, a non-Euclidean readout of the variations of the response according to some \emph{readout metric matrix} $R$, implies that $D^2 = \Delta \vect{y}^\top \cdot R \cdot \Delta \vect{y}$. With $R$ symmetric it can be diagonalized by orthonormal transforms: $R = B \cdot \lambda \cdot B^\top$. 
Therefore, this non-Eucidean metric in the response domain is equivalent to including an extra transform to a domain $\vect{y'} = \lambda^{\frac{1}{2}} \cdot B^\top \cdot \vect{y}$ where the readout is Euclidean. 
The role of this extra transform (for the metric matrix) is similar in spirit to a whitening transform (for the covariance matrix).
In this new domain, the covariance of the inner noise would be $\Sigma_{\mathcal{I}}^{y'} = \lambda^{\frac{1}{2}} \cdot B \cdot \Sigma_{\mathcal{I}} \cdot B^\top \cdot \lambda^{\frac{1}{2}}$. Therefore, Eq.~\ref{maha_in_response} can be generalized to: 
\begin{equation}
      D_{\textrm{th}}^2 = \Delta S^\top \cdot \left( \lambda^{\frac{1}{2}} \cdot B \cdot \Sigma_{\mathcal{I}} \cdot B^\top \cdot \lambda^{\frac{1}{2}} + k^2 \sigma^2 I \right)^{-1} \cdot \Delta S 
       \label{maha_in_response2}
\end{equation}
where $\Sigma_{\mathcal{I}}$ has the same expression as the one in Eq.~\ref{metric_noise1}. 
As a consequence, all the variables can be found through optimization in the same way, and all the discussion about the role of the external noise is the same: even if both factors (non-Euclidean readout and noise in the decision) are taken together, the scale of the late noise and the decision noise would remain unknown if there is no external noise at the input.

\end{document}